\documentclass[%
reprint,
 amsmath,amssymb,
 aps,
]{revtex4-2} 
\usepackage{graphicx}
\usepackage{dcolumn} 
\usepackage{bm}

\begin{document}


\title{\LARGE Reconfigurable miniaturized computational spectrometer enabled by photoelastic effect}

\author{Linjun Zhai\textsuperscript{1}}
\author{Baolei Liu\textsuperscript{2}}
 \email{liubaolei@buaa.edu.cn}
\author{Muchen Zhu\textsuperscript{1}}
\author{Yao Wang\textsuperscript{1}}
\author{Chaohao Chen\textsuperscript{3}}
\author{Zhaohua Yang\textsuperscript{2}}
\author{Lan Fu\textsuperscript{4}}
\author{Fan Wang\textsuperscript{1}}
 \email{fanwang@buaa.edu.cn}
 
\affiliation{%
 \textsuperscript{1}School of Physics, Beihang University, Beijing, 100191, China\\
\textsuperscript{2}School of Instrumentation and Optoelectronics Engineering, Beihang University, Beijing 100191, China\\
\textsuperscript{3}School of Mathematical and Physical Sciences, Faculty of Science, University of Technology Sydney, NSW 2007, Australia\\
\textsuperscript{4}\mbox{Australian Research Council Centre of Excellence for Transformative Meta-Optical Systems,} \mbox{Department of Electronic Materials Engineering,Research School of Physics,} \mbox{The Australian National University, Canberra, ACT 2600, Australia}\\}%
\date{August 8th, 2025}

\begin{abstract}
Miniatured computational spectrometers, distinguished by their compact size and lightweight, have shown great promise for on-chip and portable applications in the fields of healthcare, environmental monitoring, food safety, and industrial process monitoring. However, the common miniaturization strategies predominantly rely on advanced micro-nano fabrication and complex material engineering, limiting their scalability and affordability. Here, we present a broadband miniaturized computational spectrometer (ElastoSpec) by leveraging the photoelastic effect for easy-to-prepare and reconfigurable implementations. A single computational photoelastic spectral filter, with only two polarizers and a plastic sheet, is designed to be integrated onto the top of a CMOS sensor for snapshot spectral acquisition. The different spectral modulation units are directly generated from different spatial locations of the filter, due to the photoelastic induced chromatic polarization effect of the plastic sheet. We experimentally demonstrate that ElastoSpec offers excellent reconstruction accuracy for the measurement of both simple narrowband and complex spectra. It achieves a full width at half maximum (FWHM) error of approximately 0.2 nm for monochromatic inputs, and maintains a mean squared error (MSE) value on the order of $10^{-3}$ with only 10 spectral modulation units. Furthermore, we develop a reconfigurable strategy for enhanced spectra sensing performance through the flexibility in optimizing the modulation effectiveness and the number of spectral modulation units. This work avoids the need for complex micro-nano fabrication and specialized materials for the design of computational spectrometers, thus paving the way for the development of simple, cost-effective, and scalable solutions for on-chip and portable spectral sensing devices.
\end{abstract}

\maketitle


\section{Introduction} 

Spectrometer has been widely used in industrial inspections, clinical diagnosis, and environmental monitoring, due to its non-contact, non-destructive, real-time, and high sensitivity\cite{1,2,3,4}. Conventional spectrometers are usually equipped with bulky optical components and mechanical structures that consist of prisms or gratings, and optical systems with long focal lengths \cite{5}. This bulky, non-portable, and high-cost design cannot meet the increasingly complex and diverse needs of a wide range of applications, such as portable detection, real-time monitoring, and embedded system integration \cite{6}. Therefore, developing miniaturized, low-powered, and low-cost spectrometers is highly desirable.
Recently, the rapid progress in micro-nano fabrication and computational imaging technologies prompted the development of miniaturized computational spectrometers. Various spectral modulation materials and devices have been investigated for use in computational spectrometers, such as quantum dots \cite{7,8,9}, photonic crystals \cite{10}, liquid crystals \cite{11}, thin films \cite{12,13,14}, nanowires \cite{15,16}, and metasurfaces \cite{2,17,18}. By incorporating external electrically tunable response functions of photodetectors, computational spectrometers have also been demonstrated with simplified hardware implementations and lower system cost \cite{19,20,21,22,23}. Recently, we developed a computational on-chip spectrometer for achieving tunable spectral modulation with reduced spectral modulation units by combining the birefringence-induced chromatic polarization effect and electrochromic modulation \cite{24}. This provides a promising method for tunability and structural simplification in spectral modulation. Subsequently, a stress-engineered ultra-broadband spectrometer was reported where dispersion elements were fabricated based on the shape memory polymers \cite{25}. By engineering internal stress through temperature-controlled mechanical stretching, efficient and controllable spectral coding can be fabricated to provide a cost-effective solution for compact spectrometers. In general, the exploration of micro-nano structures and materials contributes significantly to the development of computational spectrometers for minimized size, enhanced flexibility, and improved spectral resolution. However, most of these existing techniques require precise fabrication processes and advanced micro- and nanofabrication, posing challenges in manufacturing miniaturized spectrometers for affordable and scalable applications.

In this work, we propose an on-chip computational spectrometer (ElastoSpec) based on the photoelastic effect, using an easy-to-prepare photoelastic spectral filter to enable miniaturized and reconfigurable implementations. The photoelastic spectral filter, composed of two polarizers and an embedded stress-induced plastic sheet, is directly integrated with a complementary metal oxide semiconductor (CMOS) sensor, eliminating the need for nanofabrication and complex arrayed spectral filters. The performance of the designed filter is characterized by measuring the spectral response functions at different polarization angles and different spatial locations, revealing that the filter exhibits rich spectral responses to be used for spectral sensing. Spectral modulation units are realized by selecting different spatial locations of the filter. We validate the effectiveness of ElastoSpec with the input of both simple narrowband and complex spectra. It achieves a full width at half maximum (FWHM) error of ~ 0.2 nm for monochromatic lights, and ensures accurate reconstruction with an MSE on the order of $10^{-3}$ with only 10 spectral modulation units. We also test the spectrometer with bandpass spectral filters and Light Emitting Diode (LED) lights with different colors. Furthermore, we propose a reconfigurable strategy to enhance ElastoSpec’s performance by optimizing the sampling number and spatial locations of the spectral modulation units used on the filter. Our ElastoSpec concept offers a compact, easy-to-prepare, and cost-effective strategy for designing portable optical sensing and consumer devices. 

\section{PRINCIPLE}

The photoelastic effect is a phenomenon in which stress induces birefringence with the changing of refractive index in some materials \cite{26,27}. This phenomenon is commonly observed in plastic materials, where internal stress creates variations in refractive indices along different directions. When the plastic is observed between crossed polarizers, a pattern of color can be observed. Such an effect allows the pattern to encode spectral information in different spatial locations, which can be exploited as dispersive elements for computational spectrometers \cite{24}. Figure. 1 illustrates the concept and principle of ElastoSpec. The schematic diagram of ElastoSpec is illustrated in Fig. 1(a), where the photoelastic spectral filter is composed of two polarizers and an embedded birefringent plastic sheet. Figure 1(b) illustrates the prototype photograph of ElastoSpec, in which the photoelastic spectral filter is integrated directly in front of a CMOS sensor. The compact spectrometer features a footprint of 27 mm × 27 mm and a weight of 4.91 g. As illustrated in Fig. 1(a), the reflected light field from the natural object, containing spectral information, is directed collected to illuminate the spectrometer. This reflected light passes through the first polarizer to produce linearly polarized light. The light is split into an ordinary ray and an extraordinary ray with different refractive indices after passing through the birefringent plastic sheet. The birefringence-induced phase difference between these two rays varies with spatial locations and wavelengths. Due to the non-uniform stress distribution within the plastic sheet, the modulation that varies across different spatial locations and wavelengths can form into different spectral modulation units. The intensity distribution of modulated light field behind the second polarizer is finally captured by the CMOS sensor, as shown in the following formula:
\begin{equation}
    I_{i}=\int_{\lambda_{1}}^{\lambda_{2}}F(\lambda)R_{i}(\lambda)d\lambda
\end{equation}
where i is the index of the selected spectral modulation units, and $I_i$ represents the measured light intensity corresponding to the i th modulation unit. $\lambda_1$ and $\lambda_2$  indicate the minimum and maximum wavelengths of the spectrometer's working range, respectively, and $F(\lambda)$ denotes the spectrum of arbitrary incidence.$ R_i (\lambda)$ is the spectral response function of the i th spectral modulation unit. The right part of Fig.1(b) presents an illustrative example of spectral modulation units selected from different spatial positions on the filter, while Fig.1(c) presents the normalized spectral response functions of 30 spectral modulation units of the filter. These spectral response functions work as modulation masks to encode the input spectrum $F(\lambda)$ into 30 light intensity values. The decoding algorithms, such as compressive sensing algorithms, are introduced to reconstruct the input spectrum, as illustrated in Fig. 1(d). A physical model $Y=Ax$ is formulated, in which x denotes the discrete representation of the target spectrum, A is the measurement matrix consisting of precalibrated spectral response functions, and Y is the vector of measured light intensity values. The target spectrum x is recovered by solving the linear problem under the sparsity hypothesis with the compressive sensing method \cite{28,29,30}.

\begin{figure*}
\includegraphics[width=\textwidth]{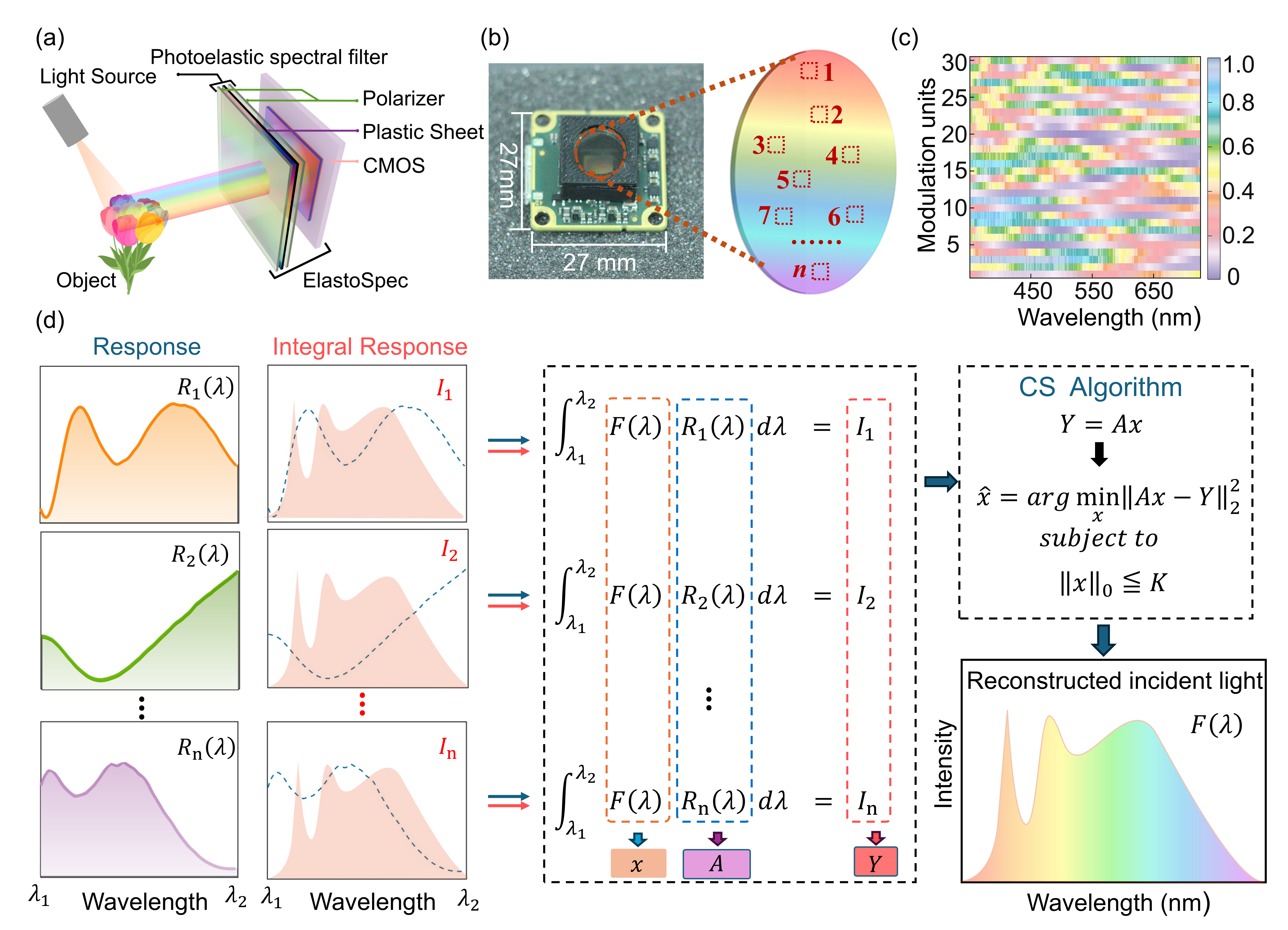}%
\caption{Scheme and principle of the proposed ElastoSpec. (a) Illustration of the ElastoSpec, in which a photoelastic spectral filter is integrated on the front of a CMOS sensor. (b) Photograph of the designed spectrometer prototype. The right part of (b) illustrates an example of spatial positions on the filter used as spectral modulation units. (c) Normalized spectral responses of different spectral modulation units, corresponding to different spatial locations of the photoelastic spectral filter indicated in (b). (d) Workflow of the designed spectrometer. The measured intensities, corresponding to different spectral modulation units, extracted from the snapshot raw image, are used to reconstruct the spectrum of incident light.}
\label{fig:1}
\end{figure*}

\begin{figure*}
\includegraphics[width=\textwidth]{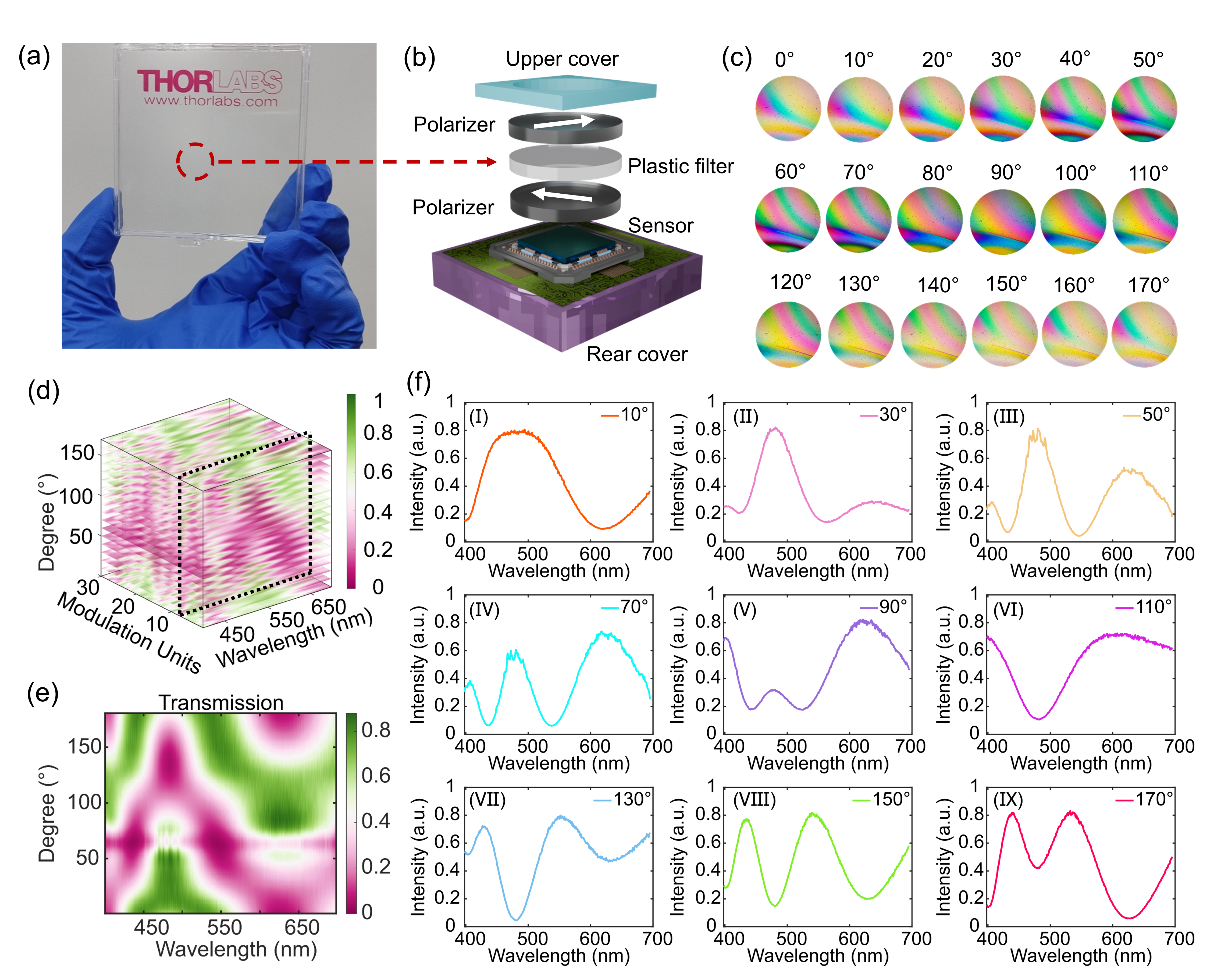}%
\caption{Optical characteristics of the photoelastic spectral filter with different polarization angles and spatial locations. (a) Photograph of the selected plastic sheet. (b) Schematic diagram of the ElastoSpec. (c) Colorful photographs of the photoelastic spectral filter at different polarization angles of the second polarizer, with the angle of the first polarizer fixed. (d) Three-dimensional representation of the transmission spectra, with different polarization angles (Polarizer 2), and spatial locations (indexed by filter number of 0, 1, ..., 30). (e) An example of transmission spectra with different polarization angles, corresponding to selected spatial locations. (f) Example spectral response curves, selected from (e), correspond to different polarization angles.}
\label{fig:2}
\end{figure*}

\begin{figure*}
\includegraphics[width=17.5cm]{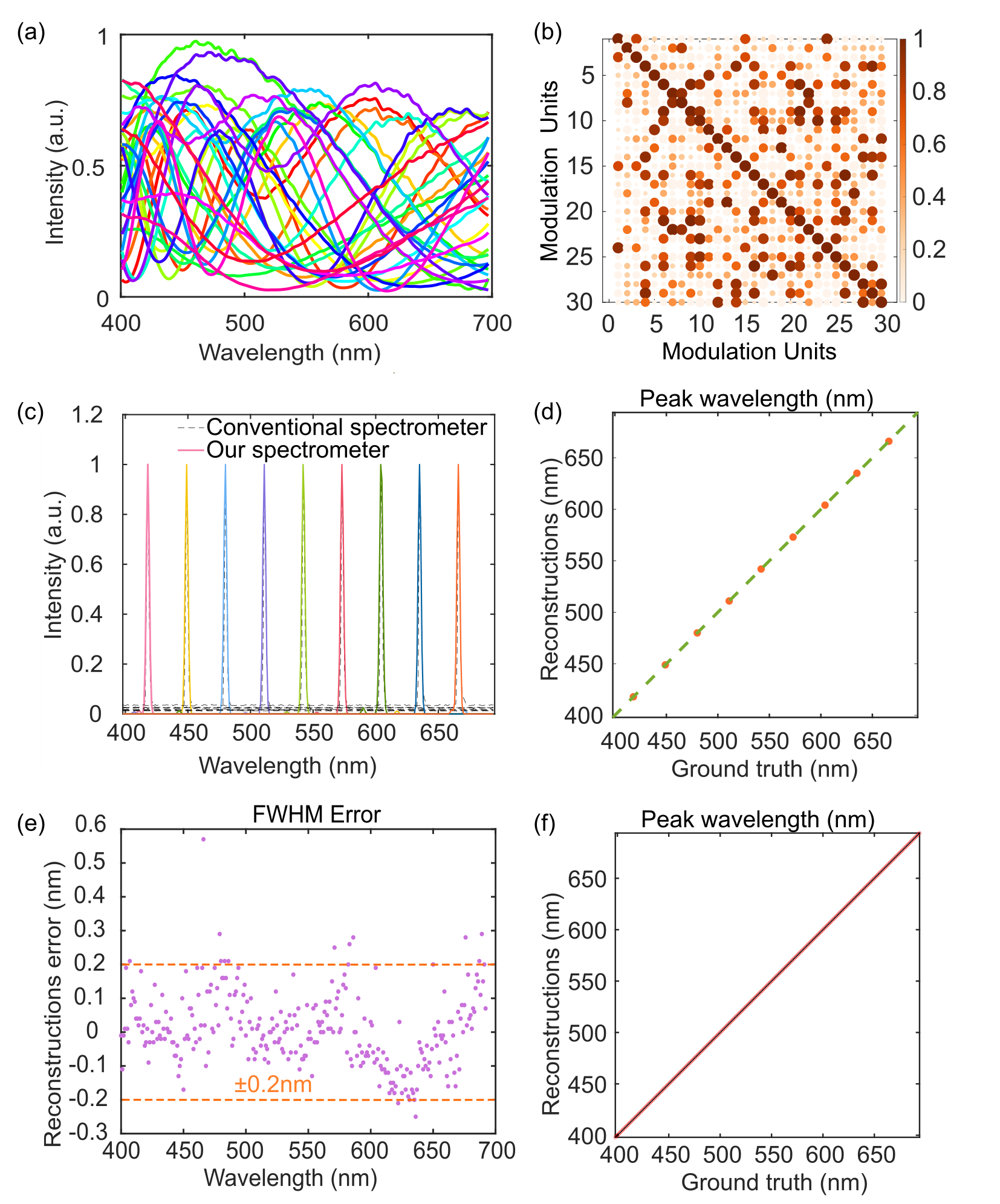}%
\caption{Spectral reconstruction results of monochromatic light. (a) Spectra response curves of the 30 spectral modulation units. (b) Correlation coefficient matrix corresponds to pairwise comparisons of the spectral response of (a). (c) Spectral reconstructions of 9 example monochromatic lights at different wavelengths in the range of 400-700 nm, compared with reference spectra measured by a commercial spectrometer. (d) Reconstruction errors of the peak wavelengths, corresponding to the reconstructed spectra in (c). (e)-(f) Reconstruction errors of FWHM and the peak wavelengths for monochromatic light in the range of 400-700 nm, with an interval of 1 nm.}
\label{fig:3}
\end{figure*}

\begin{figure*}
\includegraphics[width=15cm]{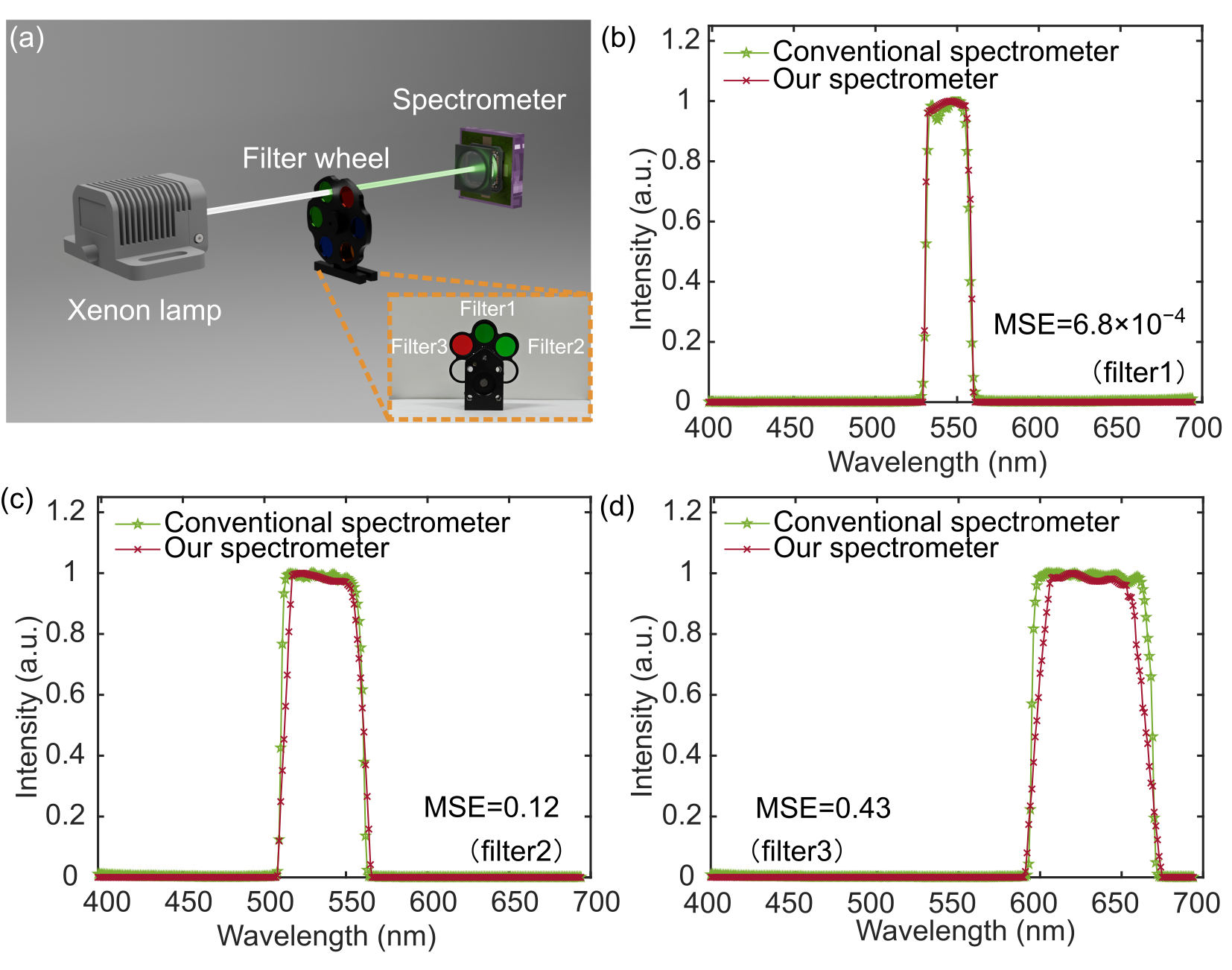}%
\caption{Reconstruction results of spectra with varying bandwidths. (a) Experimental Setup for the test of bandpass filters. A Xenon lamp (Zolix, GLORIA-X150A) is used as the white light source, and the bandpass filters are mounted on a filter wheel. (b) Spectral reconstruction of Filter 1 (LBTEK, MBF10-550-25), with the central wavelength of 550 nm and the bandwidth of 25 nm. (c) Spectral reconstruction of Filter 2 (Chroma, ET530/50M), with the central wavelength of 530 nm and the bandwidth of 50 nm. (d) Spectral reconstruction of Filter 3 (Chroma, ET630/75M), with the central wavelength of 630 nm and the bandwidth of 75 nm.}
\label{fig:4}
\end{figure*}

\begin{figure*}
\includegraphics[width=15cm]{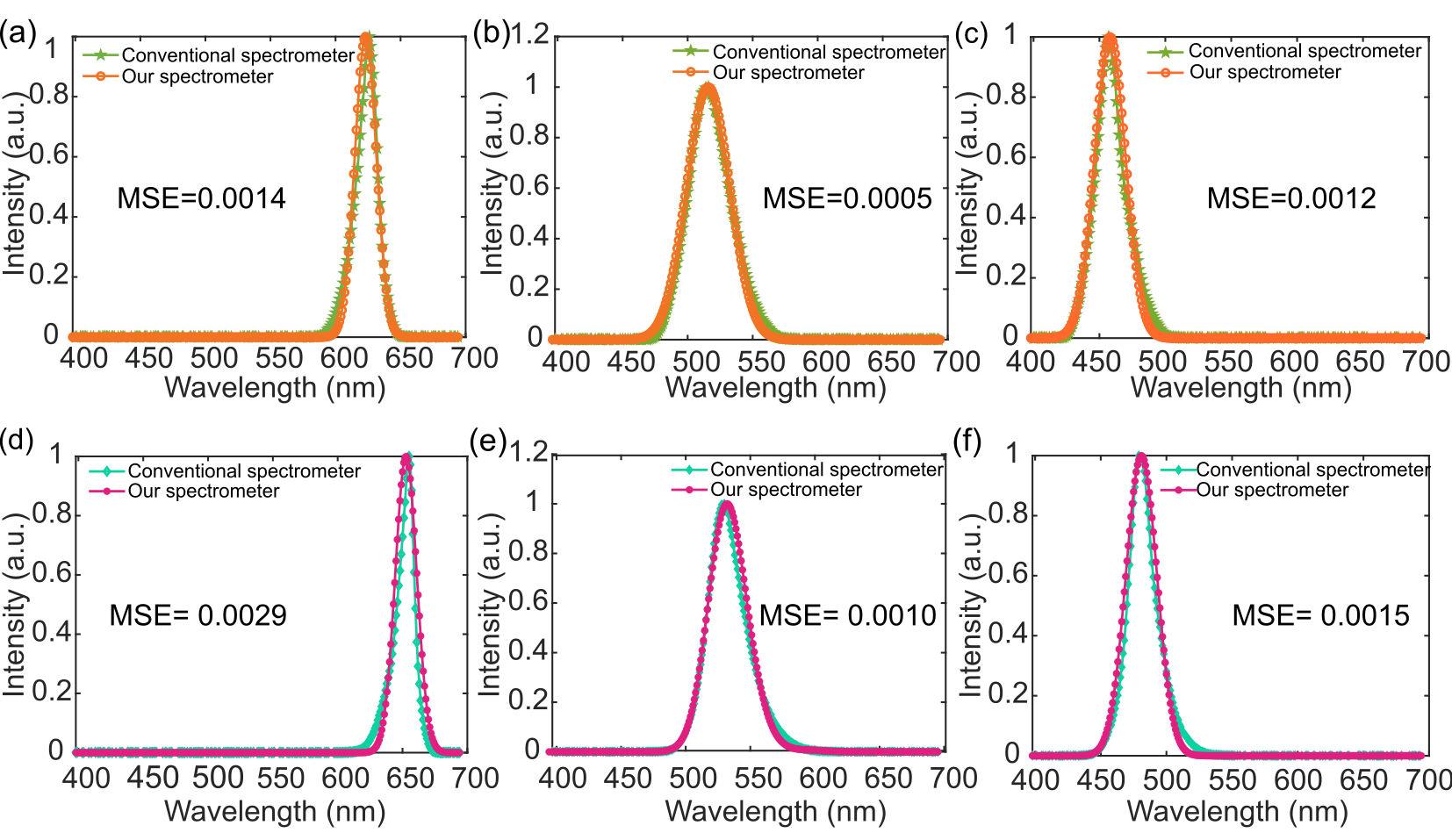}%
\caption{Spectral reconstruction results of LED light with different colors. (a)-(c) Reconstructed and measured spectra of red, green, and blue LEDs (DHC, GCI-060401, GCI-060403, GCI-060404) with central wavelengths of 620 nm, 528 nm, and 451 nm, respectively. (d)-(f) Reconstructed and measured spectra of another group of LEDs (JCOPTIX, LEFC660, LEFC530, LEFC470) with central wavelengths of 660 nm, 570 nm, and 470 nm, respectively.}
\label{fig:5}
\end{figure*}

\begin{figure*}
\includegraphics[width=15cm]{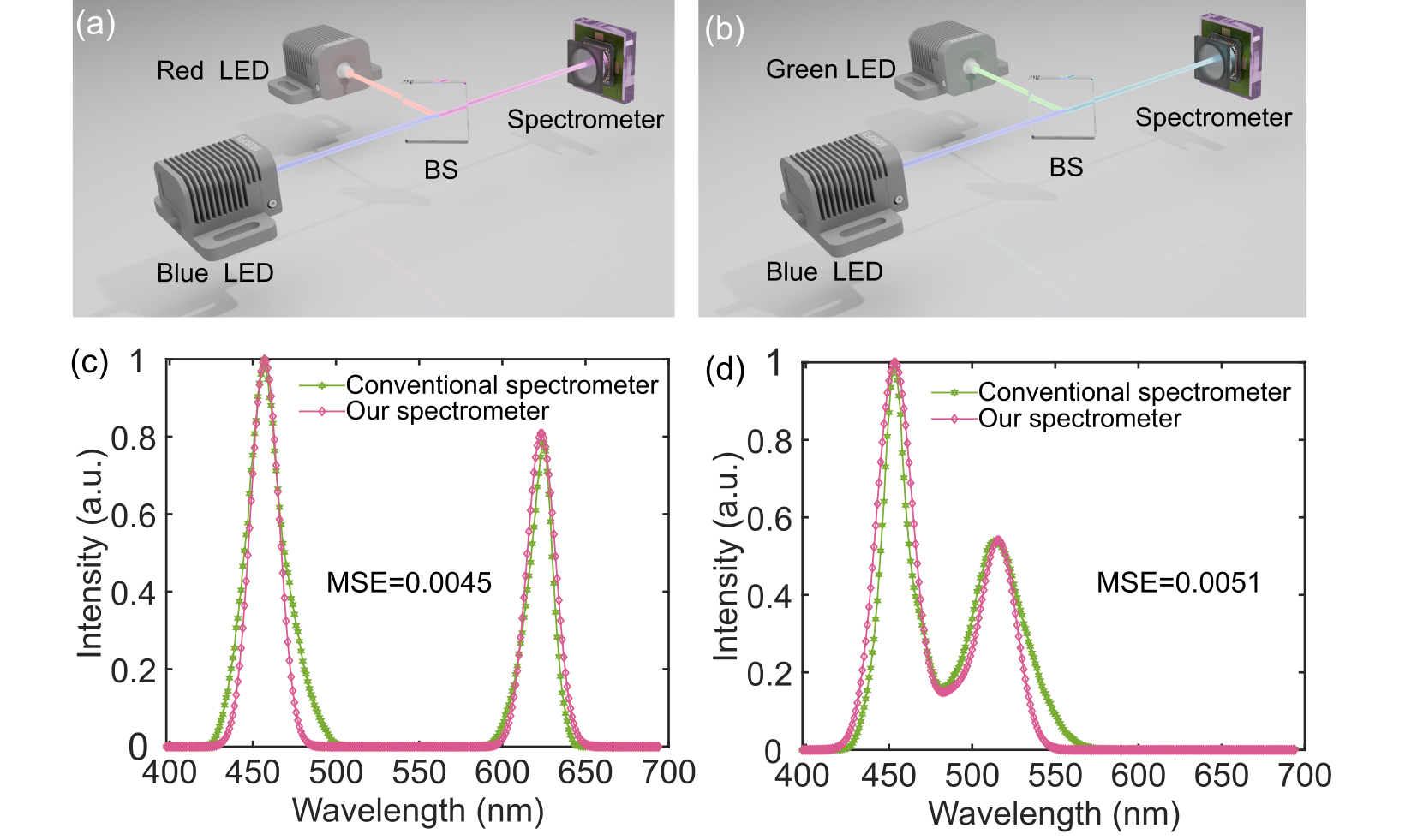}%
\caption{Spectral reconstruction results for the input double-peaked spectra. (a)-(b) Experimental setups for generating the double-peaked spectra, by combining two distinct LED sources with a beam splitter (BS). (c)-(d) Reconstructed spectra compared with reference spectra measured by a commercial spectrometer, corresponding to the input spectra in (a) and (b), respectively.}
\label{fig:6}
\end{figure*}

\begin{figure*}
\includegraphics[width=15cm]{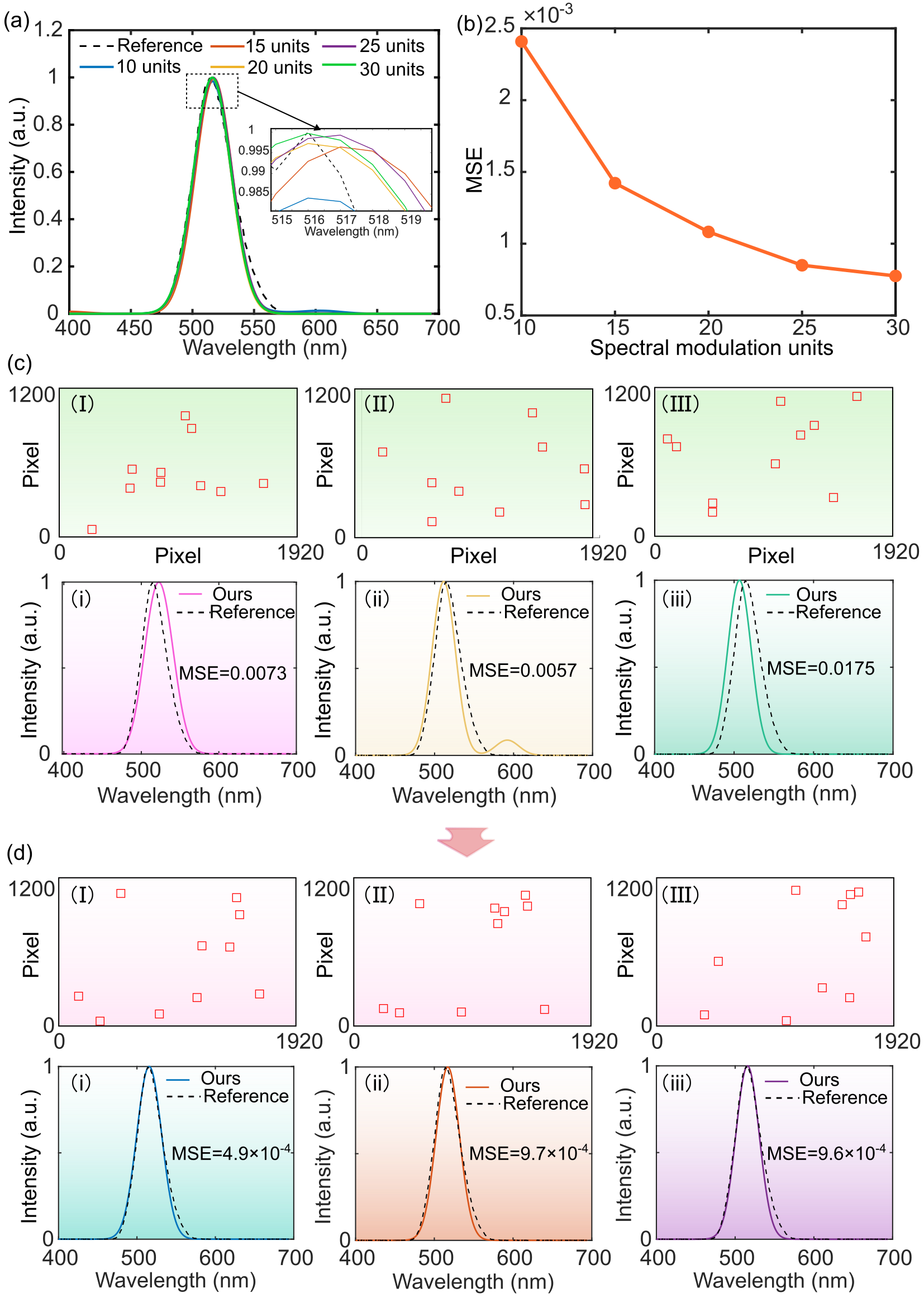}%
\caption{Proposed reconfigurable strategy to further enhance the reconstruction quality. (a) Spectral reconstruction results obtained by improving the sampling number of the spectral modulation units. (b) MSE values for the spectral reconstructions in (a). (c) Spectral results reconstructed by randomly selecting the spectral modulation units. (\uppercase\expandafter{\romannumeral1})-(\uppercase\expandafter{\romannumeral3}) present the randomly selected spatial distributions of modulation units. (\expandafter{\romannumeral1})-(\expandafter{\romannumeral3}) show the spectral reconstruction results corresponding to (\uppercase\expandafter{\romannumeral1})-(\uppercase\expandafter{\romannumeral3}) in (c). (d) Spectral results reconstructed by optimally selecting the spectral modulation units. (\uppercase\expandafter{\romannumeral1})-(\uppercase\expandafter{\romannumeral3}) present the optimally selected spatial distributions of modulation units. (\expandafter{\romannumeral1})-(\expandafter{\romannumeral3}) show the spectral reconstruction results corresponding to (\uppercase\expandafter{\romannumeral1})-(\uppercase\expandafter{\romannumeral3}) in (d).}
\label{fig:7}
\end{figure*}

\section{RESULTS AND DISCUSSION}
\subsection{Fabrication and characterization of the photoelastic spectral filter}

Figure 2 illustrates the optical response characterization of the photoelastic spectral filter at different polarization angles and spatial locations. As shown in Fig. 2(a), the used plastic sheet is cut from a commonly available optic storage box (Thorlabs, BX03), serving as the birefringent material here. The plastic sheet and two polarizers with a rotation angle (here we use 60°) construct the photoelastic spectral filter, which is then integrated on top of the CMOS sensor (daA2500-14 $\mu m$, Basler) to form ElastoSpec as illustrated in Fig. 2(b). Figure 2(c) presents the capture transmission images of the photoelastic spectral filter with various polarization angles (the second polarizer varied from 0° to 170° with an increment of 10°). It can be noted that these images under different polarization angles exhibit different color patterns. This phenomenon is attributed to the angle between the polarized light and the spatially changed refractive index of the plastic sheet, due to the photoelastic effect. We also present the three-dimensional visualization of the transmission spectra from different spectral modulation units under the different polarization angles, as shown in Fig. 2(d). These spectra can be used to select the spatial locations of spectral modulation units or choose appropriate polarization angles. An example of the polarization-dependent transmission spectra in a selected modulation unit is shown in Fig. 2(e), which corresponds to the 8th spectral modulation unit highlighted by the dashed box in Fig.2(d). This example clearly illustrates the dependence of spectral response on the polarization angles. For instance, higher transmission intensities are achieved in the range of 480-510 nm when the polarization angle is between 0° and 80°, while higher transmission intensities occur when the polarization angle is set between 80° and 120° in the range of 600-650 nm. Figure. 2(f) further shows the spectral response curves under different polarization angles of the 8th spectral modulation unit. It can be observed that a relatively high transmission is achieved under different polarization angles, which enhances optical throughput for high-efficiency spectral sensing. Moreover, the photoelastic spectral filter responds sensitively to variations of polarization angle, offering the potential to further improve diverse spectral response and sampling number by introducing tuned polarization angles as an additional modulation dimension in future works.

\subsection{Reconstruction performance of ElastoSpec}

To test the reconstruction performance of our spectrometer, we first analyzed the spectral reconstruction of the spectrometer on the spectra produced by monochromatic light. Figure 3(a) shows the experimental spectral response functions of 30 selected spectral modulation units. The pairwise correlation coefficient matrix (Fig. 3b) of these 30 modulation units yields an average off-diagonal value of 0.439, indicating low mutual correlation among the spectral responses. It is noted that we can further reduce the correlation coefficients by reconfiguring the spatial locations of the modulation units for low-correlated spectral responses. The reconstructed spectra of monochromatic lights at 9 different wavelengths, with a step of 50 nm, are presented in Fig. 3(c), with comparisons measured by a commercial spectrometer (Ocean Optics, USB4000). The maximum peak-wavelength values of the reconstructed spectra match that for reference spectra by forming a linear diagonal line (Fig. 3d). To further quantify the measurement errors, we analyze the errors of maximum peak-wavelength values and the full width at half maximum (FWHM) in the range of 400-700 nm with a step of 1 nm, as presented in Figs. 3(e)-(f). The analysis reveals that the FWHM errors at most of the wavelengths are within the range of ±0.2 nm, indicating good reconstruction accuracy in its working range. For Fig. 3(f), the results reveal that most of the maximum peak-wavelength values are tightly distributed on the diagonal line, which exhibits low maximum peak wavelength errors.

We further test the proposed spectrometer using bandpass filters with different bandwidths, as illustrated in Fig. 4. As illustrated in Fig. 4(a), the light from the Xeon lamp (Zolix, GLORIA-X150A) passes through filters with various bandwidths, which are mounted on a filter wheel, to produce spectra with different bandwidths. The used bandpass filters have the central wavelengths of 550 nm, 530 nm, and 630 nm, and the bandwidths of 25 nm, 50 nm, and 75 nm, respectively. The comparisons between the reconstructed spectra and the reference spectra measured by a commercial spectrometer are presented in Figs. 4(b)-(d). The results demonstrate that the reconstructed spectra exhibit high consistency with the reference spectra for the case of Filter 1 with the bandwidth of 25 nm, where the MSE value reaches as low as $6.8\times10^{-4}$. The fidelity shows a trend of decreasing with the increase of bandwidth, especially for the edge profiles, in which the MSE value rises to 0.12 for Filter 2 with a bandwidth of 50 nm and to 0.43 for Filter 3 with a bandwidth of 75 nm. The reason is that, under identical spectral response functions and sampling numbers, a broader spectral bandwidth corresponds to lower sparsity, thereby degrading the reconstruction quality. Nevertheless, the reconstructed results generally agree well with the commercial measurements, demonstrating good stability and reliable reconstructions.

Figure 5 presents the spectral reconstruction results with the input light sources of different LEDs. The comparisons between the reconstructed spectra and the reference spectra are presented in Figs. 5(a)-(c), with the red, green, and blue LEDs (DHC, GCI-060401, GCI-060403, GCI-060404) that have the center wavelengths of 620 nm, 528 nm, and 451 nm, respectively. The reconstructed spectra show good agreement with the reference ones in terms of both peak position and FWHM, with MSE values of 0.0014, 0.0005, and 0.0012 for the red, green, and blue LEDs, respectively. The comparison results with another group of LEDs (JCOPTIX, LEFC660, LEFC530, and LEFC470, centered at 660 nm, 530 nm, and 470 nm, respectively) are presented in Figs. 5(d)–(f), with MSE values of 0.0029, 0.0010, and 0.0015, respectively. We further investigate the correlation coefficients of the spectral response across different wavelength bands. The reconstruction of the green light spectrum performs the best, with a relatively low MSE value (0.0005) and an averaged correlation coefficient of 0.62 (420 nm-500 nm), followed by the blue light spectrum. The reconstruction accuracy for the red light spectrum is slightly lower, as indicated by a relatively high MSE value (0.0029) and an averaged correlation coefficient of 0.88 (620 nm-680 nm). This verifies that lower correlation coefficients of the spectral response functions lead to higher reconstruction quality.

To further validate its performance, we test ElastoSpec with broadband input spectra that have asymmetric profiles, as shown in Fig. 6. Figures 6(a)-(b) display the experimental setups that generate the double-peaked spectra, with varying peak spacing, by combining two different LED sources via a beam splitter. The first input spectra were generated with a red LED (DHC, GCI-060401) and a blue LED (DHC, GCI-060404), while the second one was generated with a green LED (DHC, GCI-060403) and a blue LED (DHC, GCI-060404). The reconstructed spectra are compared with the reference measured using a commercial spectrometer (Ocean Optics, USB4000), as shown in Figs. 6(c)-(d). It is noted that the spectrometer can generally reconstruct the double-peak spectra, including both the peak positions and the spatial profiles, with MSE values of 0.0045 for Fig. 6(c) and 0.0051 for Fig. 6(d), respectively. However, the reconstructed spectra in the range of 420-510 nm and 525-575 nm do not have a good match with the reference spectra, as shown in Figs. 6(c)-(d), respectively. 

To further improve the reconstruction accuracy, we explored a reconfigurable strategy for ElastoSpec, as illustrated in Fig. 7. As mentioned above, the spectral modulation units of ElastoSpec are selected from different spatial locations of the photoelastic spectral filter, corresponding to different spectral responses. Therefore, we can reconfigure the utilized spectral modulation units to enhance the reconstruction accuracy by optimizing the spatial locations of the modulation units for lower correlation coefficients of their response functions, or by simply increasing the number of these modulation units. We conducted additional experiments to evaluate the effectiveness of this reconfiguration strategy (Fig. 7). We first studied the influence of the sampling number in Figs. 7(a)-(b). Figure 7(a) compares the reconstructions with varying sampling numbers of modulation units (10, 15, 20, ..., 30). As the number increases from 10 to 30, the reconstructed spectra progressively approach the reference spectrum. The corresponding MSE values of Fig. 7(a) are shown in Fig. 7(b), indicating the improved reconstruction quality as the number of modulation units increases. Besides, a relatively low MSE value ($\approx 1.0 × 10^{-3}$) is already achieved with 20 modulation units. Further increase of the sampling number only results in limited improvement in spectral reconstruction accuracy, while potentially introducing higher computational cost. Therefore, the reconfigurable strategy suggests a trade-off between sampling number and spectral accuracy, suggesting that an appropriate sampling number is adequate to ensure reliable spectral reconstruction while with relatively low computational complexity.

Then, we studied the influence of the spatial locations of the modulation units at the photoelastic spectral filter in Figs. 7(c)-(d). Figure 7(c) compares the spectral reconstruction results obtained by randomly selected modulation units, in which three examples are presented. It can be observed that the peak positions and spectral profiles of the reconstructed spectra deviate from the reference and even exhibit secondary noisy peaks, such as Fig. 7c(\expandafter{\romannumeral2}). Figure 7(d) compares the spectral results reconstructed by selecting the optimal modulation units. It can be seen that the optimization of modulation units enables improved accuracy, with peak positions and spectral profiles closely matching the reference, and without noisy peaks. These results demonstrate that different spatial location distributions lead to variations in spectral reconstruction accuracy, indicating that the reconfigurable strategy can provide an optimized spatial location distribution of the modulation units in the calibration for better reconstruction. The above analysis demonstrates that spectral reconstruction quality depends on the configuration of spectral modulation units. A well-designed configuration balances sampling efficiency and reconstruction accuracy while optimizing the spatial arrangement of modulation units to enable high-accuracy, computationally efficient reconstruction. However, identifying such an optimized reconfigurable strategy requires more advanced algorithms, due to the large number of possible configurations in both the number and spatial location distribution of the modulation units.

\section{CONCLUSIONS}
In summary, we propose and demonstrate a low-cost, easy-to-prepare miniaturized computational spectrometer (ElastoSpec) that utilizes a photoelastic spectral filter, achieving compact and reconfigurable implementations without relying on material synthesis and micro-nano fabrication techniques that are commonly required in existing miniatured computational spectrometers. The experimental results demonstrate that the spectrometer achieves good reconstruction accuracy for both simple narrowband spectra and complex spectra, such as broadband spectra, multi-color LED spectra, and double-peaked spectra. In particular, for monochromatic spectra, the spectrometer attains a FWHM error of approximately 0.2 nm, and for complex spectra, it maintains an MSE on the order of $10^{-3}$ with only 10 spectral modulation units. Moreover, the proposed spectrometer demonstrates flexible reconfigurability, allowing adaptive tuning of both the number and spatial distribution of spectral modulation units to meet various application requirements, thereby achieving high-quality spectral reconstruction with an appropriate sampling number that balances accuracy and computational efficiency.

Attributed to its compact, portable, cost-effective, easy-to-prepare, and reconfigurable characteristics, the proposed spectrometer could be used for real-time portable spectroscopy and exhibits promising potential for integration into wearable devices, consumer electronics, and small unmanned aerial vehicles for applications in precision agriculture, disaster monitoring, health tracking, and smart home systems \cite{31,32,33}. Nonetheless, the proposed ElastoSpec still needs to be further optimized. At the present stage, the selection strategy for the numbers and spatial location distributions of the modulation units is determined manually. In the future, an optimization model will be developed to quantitatively evaluate how the numbers and spatial distributions of modulation units affect reconstruction quality, thereby facilitating the automatic design and configuration of spectral modulation units. Furthermore, the performance of ElastoSpec can be further enhanced by integrating it with advanced spectral encoding methods \cite{34,35,36} and deep learning reconstruction algorithms \cite{37,38}. Such integration also paves the way for its application in compressive imaging \cite{39,40,41} for computational hyperspectral imaging \cite{24,42}, facilitating the development of computational spectrometers in frontier fields such as biomedical, precision medicine, and space observation.

\section*{Funding}
This work was supported by the National Natural Science Foundation of China (U23A20481, 62275010), the Fundamental Research Funds for the Central Universities (KG16-3549-01), Australian Research Council Discovery Projects (DP220101417), and Discovery Early Career Researcher Awards (DE250100406).

\section*{Disclosures}
The authors declare that they have no competing interests.

\bibliography{ref}

\begin{thebibliography}{42}%
\makeatletter
\providecommand \@ifxundefined [1]{%
 \@ifx{#1\undefined}
}%
\providecommand \@ifnum [1]{%
 \ifnum #1\expandafter \@firstoftwo
 \else \expandafter \@secondoftwo
 \fi
}%
\providecommand \@ifx [1]{%
 \ifx #1\expandafter \@firstoftwo
 \else \expandafter \@secondoftwo
 \fi
}%
\providecommand \natexlab [1]{#1}%
\providecommand \enquote  [1]{``#1''}%
\providecommand \bibnamefont  [1]{#1}%
\providecommand \bibfnamefont [1]{#1}%
\providecommand \citenamefont [1]{#1}%
\providecommand \href@noop [0]{\@secondoftwo}%
\providecommand \href [0]{\begingroup \@sanitize@url \@href}%
\providecommand \@href[1]{\@@startlink{#1}\@@href}%
\providecommand \@@href[1]{\endgroup#1\@@endlink}%
\providecommand \@sanitize@url [0]{\catcode `\\12\catcode `\$12\catcode `\&12\catcode `\#12\catcode `\^12\catcode `\_12\catcode `\%12\relax}%
\providecommand \@@startlink[1]{}%
\providecommand \@@endlink[0]{}%
\providecommand \url  [0]{\begingroup\@sanitize@url \@url }%
\providecommand \@url [1]{\endgroup\@href {#1}{\urlprefix }}%
\providecommand \urlprefix  [0]{URL }%
\providecommand \Eprint [0]{\href }%
\providecommand \doibase [0]{https://doi.org/}%
\providecommand \selectlanguage [0]{\@gobble}%
\providecommand \bibinfo  [0]{\@secondoftwo}%
\providecommand \bibfield  [0]{\@secondoftwo}%
\providecommand \translation [1]{[#1]}%
\providecommand \BibitemOpen [0]{}%
\providecommand \bibitemStop [0]{}%
\providecommand \bibitemNoStop [0]{.\EOS\space}%
\providecommand \EOS [0]{\spacefactor3000\relax}%
\providecommand \BibitemShut  [1]{\csname bibitem#1\endcsname}%
\let\auto@bib@innerbib\@empty
\bibitem [{\citenamefont {Yang}\ \emph {et~al.}(2021)\citenamefont {Yang}, \citenamefont {Albrow-Owen}, \citenamefont {Cai},\ and\ \citenamefont {Hasan}}]{1}%
  \BibitemOpen
  \bibfield  {author} {\bibinfo {author} {\bibfnamefont {Z.}~\bibnamefont {Yang}}, \bibinfo {author} {\bibfnamefont {T.}~\bibnamefont {Albrow-Owen}}, \bibinfo {author} {\bibfnamefont {W.}~\bibnamefont {Cai}},\ and\ \bibinfo {author} {\bibfnamefont {T.}~\bibnamefont {Hasan}},\ }\bibfield  {title} {\bibinfo {title} {Miniaturization of optical spectrometers},\ }\href@noop {} {\bibfield  {journal} {\bibinfo  {journal} {Science}\ }\textbf {\bibinfo {volume} {371}},\ \bibinfo {pages} {eabe0722} (\bibinfo {year} {2021})}\BibitemShut {NoStop}%
\bibitem [{\citenamefont {Xiong}\ \emph {et~al.}(2022)\citenamefont {Xiong}, \citenamefont {Cai}, \citenamefont {Cui}, \citenamefont {Huang}, \citenamefont {Yang}, \citenamefont {Zhu}, \citenamefont {Li}, \citenamefont {Hong}, \citenamefont {Rao}, \citenamefont {Zheng} \emph {et~al.}}]{2}%
  \BibitemOpen
  \bibfield  {author} {\bibinfo {author} {\bibfnamefont {J.}~\bibnamefont {Xiong}}, \bibinfo {author} {\bibfnamefont {X.}~\bibnamefont {Cai}}, \bibinfo {author} {\bibfnamefont {K.}~\bibnamefont {Cui}}, \bibinfo {author} {\bibfnamefont {Y.}~\bibnamefont {Huang}}, \bibinfo {author} {\bibfnamefont {J.}~\bibnamefont {Yang}}, \bibinfo {author} {\bibfnamefont {H.}~\bibnamefont {Zhu}}, \bibinfo {author} {\bibfnamefont {W.}~\bibnamefont {Li}}, \bibinfo {author} {\bibfnamefont {B.}~\bibnamefont {Hong}}, \bibinfo {author} {\bibfnamefont {S.}~\bibnamefont {Rao}}, \bibinfo {author} {\bibfnamefont {Z.}~\bibnamefont {Zheng}}, \emph {et~al.},\ }\bibfield  {title} {\bibinfo {title} {Dynamic brain spectrum acquired by a real-time ultraspectral imaging chip with reconfigurable metasurfaces},\ }\href@noop {} {\bibfield  {journal} {\bibinfo  {journal} {Optica}\ }\textbf {\bibinfo {volume} {9}},\ \bibinfo {pages} {461} (\bibinfo {year} {2022})}\BibitemShut {NoStop}%
\bibitem [{\citenamefont {Zhou}\ \emph {et~al.}(2023)\citenamefont {Zhou}, \citenamefont {Sun}, \citenamefont {Guo}, \citenamefont {Min}, \citenamefont {Wang},\ and\ \citenamefont {Li}}]{3}%
  \BibitemOpen
  \bibfield  {author} {\bibinfo {author} {\bibfnamefont {Y.}~\bibnamefont {Zhou}}, \bibinfo {author} {\bibfnamefont {H.}~\bibnamefont {Sun}}, \bibinfo {author} {\bibfnamefont {L.}~\bibnamefont {Guo}}, \bibinfo {author} {\bibfnamefont {L.}~\bibnamefont {Min}}, \bibinfo {author} {\bibfnamefont {M.}~\bibnamefont {Wang}},\ and\ \bibinfo {author} {\bibfnamefont {L.}~\bibnamefont {Li}},\ }\bibfield  {title} {\bibinfo {title} {Emerging computational micro-spectrometers—from complex system integration to simple in situ modulation},\ }\href@noop {} {\bibfield  {journal} {\bibinfo  {journal} {Small Methods}\ }\textbf {\bibinfo {volume} {7}},\ \bibinfo {pages} {2300479} (\bibinfo {year} {2023})}\BibitemShut {NoStop}%
\bibitem [{\citenamefont {Yoon}\ \emph {et~al.}(2019)\citenamefont {Yoon}, \citenamefont {Joseph}, \citenamefont {Waterhouse}, \citenamefont {Luthman}, \citenamefont {Gordon}, \citenamefont {Di~Pietro}, \citenamefont {Januszewicz}, \citenamefont {Fitzgerald},\ and\ \citenamefont {Bohndiek}}]{4}%
  \BibitemOpen
  \bibfield  {author} {\bibinfo {author} {\bibfnamefont {J.}~\bibnamefont {Yoon}}, \bibinfo {author} {\bibfnamefont {J.}~\bibnamefont {Joseph}}, \bibinfo {author} {\bibfnamefont {D.~J.}\ \bibnamefont {Waterhouse}}, \bibinfo {author} {\bibfnamefont {A.~S.}\ \bibnamefont {Luthman}}, \bibinfo {author} {\bibfnamefont {G.~S.}\ \bibnamefont {Gordon}}, \bibinfo {author} {\bibfnamefont {M.}~\bibnamefont {Di~Pietro}}, \bibinfo {author} {\bibfnamefont {W.}~\bibnamefont {Januszewicz}}, \bibinfo {author} {\bibfnamefont {R.~C.}\ \bibnamefont {Fitzgerald}},\ and\ \bibinfo {author} {\bibfnamefont {S.~E.}\ \bibnamefont {Bohndiek}},\ }\bibfield  {title} {\bibinfo {title} {A clinically translatable hyperspectral endoscopy (hyse) system for imaging the gastrointestinal tract},\ }\href@noop {} {\bibfield  {journal} {\bibinfo  {journal} {Nature communications}\ }\textbf {\bibinfo {volume} {10}},\ \bibinfo {pages} {1902} (\bibinfo {year} {2019})}\BibitemShut {NoStop}%
\bibitem [{\citenamefont {Xue}\ \emph {et~al.}(2024)\citenamefont {Xue}, \citenamefont {Yang}, \citenamefont {Ma}, \citenamefont {Zhang}, \citenamefont {Zhang}, \citenamefont {Lan}, \citenamefont {Gao}, \citenamefont {Zhang},\ and\ \citenamefont {Tang}}]{5}%
  \BibitemOpen
  \bibfield  {author} {\bibinfo {author} {\bibfnamefont {Q.}~\bibnamefont {Xue}}, \bibinfo {author} {\bibfnamefont {Y.}~\bibnamefont {Yang}}, \bibinfo {author} {\bibfnamefont {W.}~\bibnamefont {Ma}}, \bibinfo {author} {\bibfnamefont {H.}~\bibnamefont {Zhang}}, \bibinfo {author} {\bibfnamefont {D.}~\bibnamefont {Zhang}}, \bibinfo {author} {\bibfnamefont {X.}~\bibnamefont {Lan}}, \bibinfo {author} {\bibfnamefont {L.}~\bibnamefont {Gao}}, \bibinfo {author} {\bibfnamefont {J.}~\bibnamefont {Zhang}},\ and\ \bibinfo {author} {\bibfnamefont {J.}~\bibnamefont {Tang}},\ }\bibfield  {title} {\bibinfo {title} {Advances in miniaturized computational spectrometers},\ }\href@noop {} {\bibfield  {journal} {\bibinfo  {journal} {Advanced Science}\ }\textbf {\bibinfo {volume} {11}},\ \bibinfo {pages} {2404448} (\bibinfo {year} {2024})}\BibitemShut {NoStop}%
\bibitem [{\citenamefont {Li}\ \emph {et~al.}(2022)\citenamefont {Li}, \citenamefont {Yao}, \citenamefont {Xia}, \citenamefont {Wang}, \citenamefont {Cheng}, \citenamefont {Penty}, \citenamefont {Fainman},\ and\ \citenamefont {Pan}}]{6}%
  \BibitemOpen
  \bibfield  {author} {\bibinfo {author} {\bibfnamefont {A.}~\bibnamefont {Li}}, \bibinfo {author} {\bibfnamefont {C.}~\bibnamefont {Yao}}, \bibinfo {author} {\bibfnamefont {J.}~\bibnamefont {Xia}}, \bibinfo {author} {\bibfnamefont {H.}~\bibnamefont {Wang}}, \bibinfo {author} {\bibfnamefont {Q.}~\bibnamefont {Cheng}}, \bibinfo {author} {\bibfnamefont {R.}~\bibnamefont {Penty}}, \bibinfo {author} {\bibfnamefont {Y.}~\bibnamefont {Fainman}},\ and\ \bibinfo {author} {\bibfnamefont {S.}~\bibnamefont {Pan}},\ }\bibfield  {title} {\bibinfo {title} {Advances in cost-effective integrated spectrometers},\ }\href@noop {} {\bibfield  {journal} {\bibinfo  {journal} {Light: Science \& Applications}\ }\textbf {\bibinfo {volume} {11}},\ \bibinfo {pages} {174} (\bibinfo {year} {2022})}\BibitemShut {NoStop}%
\bibitem [{\citenamefont {Bao}\ and\ \citenamefont {Bawendi}(2015)}]{7}%
  \BibitemOpen
  \bibfield  {author} {\bibinfo {author} {\bibfnamefont {J.}~\bibnamefont {Bao}}\ and\ \bibinfo {author} {\bibfnamefont {M.~G.}\ \bibnamefont {Bawendi}},\ }\bibfield  {title} {\bibinfo {title} {A colloidal quantum dot spectrometer},\ }\href@noop {} {\bibfield  {journal} {\bibinfo  {journal} {Nature}\ }\textbf {\bibinfo {volume} {523}},\ \bibinfo {pages} {67} (\bibinfo {year} {2015})}\BibitemShut {NoStop}%
\bibitem [{\citenamefont {Meng}\ \emph {et~al.}(2024)\citenamefont {Meng}, \citenamefont {Gao}, \citenamefont {Wang}, \citenamefont {Li}, \citenamefont {Wang}, \citenamefont {Zhao},\ and\ \citenamefont {Sun}}]{8}%
  \BibitemOpen
  \bibfield  {author} {\bibinfo {author} {\bibfnamefont {H.}~\bibnamefont {Meng}}, \bibinfo {author} {\bibfnamefont {Y.}~\bibnamefont {Gao}}, \bibinfo {author} {\bibfnamefont {X.}~\bibnamefont {Wang}}, \bibinfo {author} {\bibfnamefont {X.}~\bibnamefont {Li}}, \bibinfo {author} {\bibfnamefont {L.}~\bibnamefont {Wang}}, \bibinfo {author} {\bibfnamefont {X.}~\bibnamefont {Zhao}},\ and\ \bibinfo {author} {\bibfnamefont {B.}~\bibnamefont {Sun}},\ }\bibfield  {title} {\bibinfo {title} {Quantum dot-enabled infrared hyperspectral imaging with single-pixel detection},\ }\href@noop {} {\bibfield  {journal} {\bibinfo  {journal} {Light: Science \& Applications}\ }\textbf {\bibinfo {volume} {13}},\ \bibinfo {pages} {121} (\bibinfo {year} {2024})}\BibitemShut {NoStop}%
\bibitem [{\citenamefont {Yin}\ \emph {et~al.}(2023)\citenamefont {Yin}, \citenamefont {Liu}, \citenamefont {Guan}, \citenamefont {Xie}, \citenamefont {Lu},\ and\ \citenamefont {Wang}}]{9}%
  \BibitemOpen
  \bibfield  {author} {\bibinfo {author} {\bibfnamefont {Z.}~\bibnamefont {Yin}}, \bibinfo {author} {\bibfnamefont {Q.}~\bibnamefont {Liu}}, \bibinfo {author} {\bibfnamefont {X.}~\bibnamefont {Guan}}, \bibinfo {author} {\bibfnamefont {M.}~\bibnamefont {Xie}}, \bibinfo {author} {\bibfnamefont {W.}~\bibnamefont {Lu}},\ and\ \bibinfo {author} {\bibfnamefont {S.}~\bibnamefont {Wang}},\ }\bibfield  {title} {\bibinfo {title} {Quantum-dot light-chip micro-spectrometer},\ }\href@noop {} {\bibfield  {journal} {\bibinfo  {journal} {Optics Letters}\ }\textbf {\bibinfo {volume} {48}},\ \bibinfo {pages} {3371} (\bibinfo {year} {2023})}\BibitemShut {NoStop}%
\bibitem [{\citenamefont {Wang}\ \emph {et~al.}(2019)\citenamefont {Wang}, \citenamefont {Yi}, \citenamefont {Chen}, \citenamefont {Zhou}, \citenamefont {Luk}, \citenamefont {James}, \citenamefont {Nogan}, \citenamefont {Ross}, \citenamefont {Joe}, \citenamefont {Shahsafi} \emph {et~al.}}]{10}%
  \BibitemOpen
  \bibfield  {author} {\bibinfo {author} {\bibfnamefont {Z.}~\bibnamefont {Wang}}, \bibinfo {author} {\bibfnamefont {S.}~\bibnamefont {Yi}}, \bibinfo {author} {\bibfnamefont {A.}~\bibnamefont {Chen}}, \bibinfo {author} {\bibfnamefont {M.}~\bibnamefont {Zhou}}, \bibinfo {author} {\bibfnamefont {T.~S.}\ \bibnamefont {Luk}}, \bibinfo {author} {\bibfnamefont {A.}~\bibnamefont {James}}, \bibinfo {author} {\bibfnamefont {J.}~\bibnamefont {Nogan}}, \bibinfo {author} {\bibfnamefont {W.}~\bibnamefont {Ross}}, \bibinfo {author} {\bibfnamefont {G.}~\bibnamefont {Joe}}, \bibinfo {author} {\bibfnamefont {A.}~\bibnamefont {Shahsafi}}, \emph {et~al.},\ }\bibfield  {title} {\bibinfo {title} {Single-shot on-chip spectral sensors based on photonic crystal slabs},\ }\href@noop {} {\bibfield  {journal} {\bibinfo  {journal} {Nature communications}\ }\textbf {\bibinfo {volume} {10}},\ \bibinfo {pages} {1020} (\bibinfo {year} {2019})}\BibitemShut {NoStop}%
\bibitem [{\citenamefont {August}\ and\ \citenamefont {Stern}(2013)}]{11}%
  \BibitemOpen
  \bibfield  {author} {\bibinfo {author} {\bibfnamefont {Y.}~\bibnamefont {August}}\ and\ \bibinfo {author} {\bibfnamefont {A.}~\bibnamefont {Stern}},\ }\bibfield  {title} {\bibinfo {title} {Compressive sensing spectrometry based on liquid crystal devices},\ }\href@noop {} {\bibfield  {journal} {\bibinfo  {journal} {Optics letters}\ }\textbf {\bibinfo {volume} {38}},\ \bibinfo {pages} {4996} (\bibinfo {year} {2013})}\BibitemShut {NoStop}%
\bibitem [{\citenamefont {Kim}\ \emph {et~al.}(2019)\citenamefont {Kim}, \citenamefont {Lee}, \citenamefont {Lee}, \citenamefont {Lee},\ and\ \citenamefont {Lee}}]{12}%
  \BibitemOpen
  \bibfield  {author} {\bibinfo {author} {\bibfnamefont {C.}~\bibnamefont {Kim}}, \bibinfo {author} {\bibfnamefont {W.-B.}\ \bibnamefont {Lee}}, \bibinfo {author} {\bibfnamefont {S.~K.}\ \bibnamefont {Lee}}, \bibinfo {author} {\bibfnamefont {Y.~T.}\ \bibnamefont {Lee}},\ and\ \bibinfo {author} {\bibfnamefont {H.-N.}\ \bibnamefont {Lee}},\ }\bibfield  {title} {\bibinfo {title} {Fabrication of 2d thin-film filter-array for compressive sensing spectroscopy},\ }\href@noop {} {\bibfield  {journal} {\bibinfo  {journal} {Optics and Lasers in Engineering}\ }\textbf {\bibinfo {volume} {115}},\ \bibinfo {pages} {53} (\bibinfo {year} {2019})}\BibitemShut {NoStop}%
\bibitem [{\citenamefont {Wen}\ \emph {et~al.}(2022)\citenamefont {Wen}, \citenamefont {Hao}, \citenamefont {Gao}, \citenamefont {Wang}, \citenamefont {Mo}, \citenamefont {Yuan}, \citenamefont {Chen}, \citenamefont {Wang}, \citenamefont {Zhang}, \citenamefont {Shao} \emph {et~al.}}]{13}%
  \BibitemOpen
  \bibfield  {author} {\bibinfo {author} {\bibfnamefont {J.}~\bibnamefont {Wen}}, \bibinfo {author} {\bibfnamefont {L.}~\bibnamefont {Hao}}, \bibinfo {author} {\bibfnamefont {C.}~\bibnamefont {Gao}}, \bibinfo {author} {\bibfnamefont {H.}~\bibnamefont {Wang}}, \bibinfo {author} {\bibfnamefont {K.}~\bibnamefont {Mo}}, \bibinfo {author} {\bibfnamefont {W.}~\bibnamefont {Yuan}}, \bibinfo {author} {\bibfnamefont {X.}~\bibnamefont {Chen}}, \bibinfo {author} {\bibfnamefont {Y.}~\bibnamefont {Wang}}, \bibinfo {author} {\bibfnamefont {Y.}~\bibnamefont {Zhang}}, \bibinfo {author} {\bibfnamefont {Y.}~\bibnamefont {Shao}}, \emph {et~al.},\ }\bibfield  {title} {\bibinfo {title} {Deep learning-based miniaturized all-dielectric ultracompact film spectrometer},\ }\href@noop {} {\bibfield  {journal} {\bibinfo  {journal} {Acs Photonics}\ }\textbf {\bibinfo {volume} {10}},\ \bibinfo {pages} {225} (\bibinfo {year} {2022})}\BibitemShut {NoStop}%
\bibitem [{\citenamefont {Bhatti}\ \emph {et~al.}(2025)\citenamefont {Bhatti}, \citenamefont {Lee}, \citenamefont {Kim}, \citenamefont {Choi}, \citenamefont {Yoon},\ and\ \citenamefont {Lee}}]{14}%
  \BibitemOpen
  \bibfield  {author} {\bibinfo {author} {\bibfnamefont {D.~S.}\ \bibnamefont {Bhatti}}, \bibinfo {author} {\bibfnamefont {J.}~\bibnamefont {Lee}}, \bibinfo {author} {\bibfnamefont {C.}~\bibnamefont {Kim}}, \bibinfo {author} {\bibfnamefont {Y.}~\bibnamefont {Choi}}, \bibinfo {author} {\bibfnamefont {H.~H.}\ \bibnamefont {Yoon}},\ and\ \bibinfo {author} {\bibfnamefont {H.-N.}\ \bibnamefont {Lee}},\ }\bibfield  {title} {\bibinfo {title} {Deep learning-based single-shot computational spectrometer using multilayer thin films},\ }\href@noop {} {\bibfield  {journal} {\bibinfo  {journal} {Scientific Reports}\ }\textbf {\bibinfo {volume} {15}},\ \bibinfo {pages} {21232} (\bibinfo {year} {2025})}\BibitemShut {NoStop}%
\bibitem [{\citenamefont {Yang}\ \emph {et~al.}(2019)\citenamefont {Yang}, \citenamefont {Albrow-Owen}, \citenamefont {Cui}, \citenamefont {Alexander-Webber}, \citenamefont {Gu}, \citenamefont {Wang}, \citenamefont {Wu}, \citenamefont {Zhuge}, \citenamefont {Williams}, \citenamefont {Wang} \emph {et~al.}}]{15}%
  \BibitemOpen
  \bibfield  {author} {\bibinfo {author} {\bibfnamefont {Z.}~\bibnamefont {Yang}}, \bibinfo {author} {\bibfnamefont {T.}~\bibnamefont {Albrow-Owen}}, \bibinfo {author} {\bibfnamefont {H.}~\bibnamefont {Cui}}, \bibinfo {author} {\bibfnamefont {J.}~\bibnamefont {Alexander-Webber}}, \bibinfo {author} {\bibfnamefont {F.}~\bibnamefont {Gu}}, \bibinfo {author} {\bibfnamefont {X.}~\bibnamefont {Wang}}, \bibinfo {author} {\bibfnamefont {T.-C.}\ \bibnamefont {Wu}}, \bibinfo {author} {\bibfnamefont {M.}~\bibnamefont {Zhuge}}, \bibinfo {author} {\bibfnamefont {C.}~\bibnamefont {Williams}}, \bibinfo {author} {\bibfnamefont {P.}~\bibnamefont {Wang}}, \emph {et~al.},\ }\bibfield  {title} {\bibinfo {title} {Single-nanowire spectrometers},\ }\href@noop {} {\bibfield  {journal} {\bibinfo  {journal} {Science}\ }\textbf {\bibinfo {volume} {365}},\ \bibinfo {pages} {1017} (\bibinfo {year} {2019})}\BibitemShut {NoStop}%
\bibitem [{\citenamefont {Meng}\ \emph {et~al.}(2019)\citenamefont {Meng}, \citenamefont {Cadusch},\ and\ \citenamefont {Crozier}}]{16}%
  \BibitemOpen
  \bibfield  {author} {\bibinfo {author} {\bibfnamefont {J.}~\bibnamefont {Meng}}, \bibinfo {author} {\bibfnamefont {J.~J.}\ \bibnamefont {Cadusch}},\ and\ \bibinfo {author} {\bibfnamefont {K.~B.}\ \bibnamefont {Crozier}},\ }\bibfield  {title} {\bibinfo {title} {Detector-only spectrometer based on structurally colored silicon nanowires and a reconstruction algorithm},\ }\href@noop {} {\bibfield  {journal} {\bibinfo  {journal} {Nano Letters}\ }\textbf {\bibinfo {volume} {20}},\ \bibinfo {pages} {320} (\bibinfo {year} {2019})}\BibitemShut {NoStop}%
\bibitem [{\citenamefont {Faraji-Dana}\ \emph {et~al.}(2018)\citenamefont {Faraji-Dana}, \citenamefont {Arbabi}, \citenamefont {Arbabi}, \citenamefont {Kamali}, \citenamefont {Kwon},\ and\ \citenamefont {Faraon}}]{17}%
  \BibitemOpen
  \bibfield  {author} {\bibinfo {author} {\bibfnamefont {M.}~\bibnamefont {Faraji-Dana}}, \bibinfo {author} {\bibfnamefont {E.}~\bibnamefont {Arbabi}}, \bibinfo {author} {\bibfnamefont {A.}~\bibnamefont {Arbabi}}, \bibinfo {author} {\bibfnamefont {S.~M.}\ \bibnamefont {Kamali}}, \bibinfo {author} {\bibfnamefont {H.}~\bibnamefont {Kwon}},\ and\ \bibinfo {author} {\bibfnamefont {A.}~\bibnamefont {Faraon}},\ }\bibfield  {title} {\bibinfo {title} {Compact folded metasurface spectrometer},\ }\href@noop {} {\bibfield  {journal} {\bibinfo  {journal} {Nature communications}\ }\textbf {\bibinfo {volume} {9}},\ \bibinfo {pages} {4196} (\bibinfo {year} {2018})}\BibitemShut {NoStop}%
\bibitem [{\citenamefont {Cai}\ \emph {et~al.}(2024)\citenamefont {Cai}, \citenamefont {Li}, \citenamefont {Zhang}, \citenamefont {Jiang}, \citenamefont {Chen}, \citenamefont {Qu}, \citenamefont {Zhang}, \citenamefont {Xiao}, \citenamefont {Han}, \citenamefont {Yu} \emph {et~al.}}]{18}%
  \BibitemOpen
  \bibfield  {author} {\bibinfo {author} {\bibfnamefont {G.}~\bibnamefont {Cai}}, \bibinfo {author} {\bibfnamefont {Y.}~\bibnamefont {Li}}, \bibinfo {author} {\bibfnamefont {Y.}~\bibnamefont {Zhang}}, \bibinfo {author} {\bibfnamefont {X.}~\bibnamefont {Jiang}}, \bibinfo {author} {\bibfnamefont {Y.}~\bibnamefont {Chen}}, \bibinfo {author} {\bibfnamefont {G.}~\bibnamefont {Qu}}, \bibinfo {author} {\bibfnamefont {X.}~\bibnamefont {Zhang}}, \bibinfo {author} {\bibfnamefont {S.}~\bibnamefont {Xiao}}, \bibinfo {author} {\bibfnamefont {J.}~\bibnamefont {Han}}, \bibinfo {author} {\bibfnamefont {S.}~\bibnamefont {Yu}}, \emph {et~al.},\ }\bibfield  {title} {\bibinfo {title} {Compact angle-resolved metasurface spectrometer},\ }\href@noop {} {\bibfield  {journal} {\bibinfo  {journal} {Nature Materials}\ }\textbf {\bibinfo {volume} {23}},\ \bibinfo {pages} {71} (\bibinfo {year} {2024})}\BibitemShut {NoStop}%
\bibitem [{\citenamefont {Wang}\ \emph {et~al.}(2024)\citenamefont {Wang}, \citenamefont {Pan}, \citenamefont {Wang}, \citenamefont {Zhang}, \citenamefont {Zhou}, \citenamefont {Yao}, \citenamefont {Wu}, \citenamefont {Ren}, \citenamefont {Wang}, \citenamefont {Ji} \emph {et~al.}}]{19}%
  \BibitemOpen
  \bibfield  {author} {\bibinfo {author} {\bibfnamefont {J.}~\bibnamefont {Wang}}, \bibinfo {author} {\bibfnamefont {B.}~\bibnamefont {Pan}}, \bibinfo {author} {\bibfnamefont {Z.}~\bibnamefont {Wang}}, \bibinfo {author} {\bibfnamefont {J.}~\bibnamefont {Zhang}}, \bibinfo {author} {\bibfnamefont {Z.}~\bibnamefont {Zhou}}, \bibinfo {author} {\bibfnamefont {L.}~\bibnamefont {Yao}}, \bibinfo {author} {\bibfnamefont {Y.}~\bibnamefont {Wu}}, \bibinfo {author} {\bibfnamefont {W.}~\bibnamefont {Ren}}, \bibinfo {author} {\bibfnamefont {J.}~\bibnamefont {Wang}}, \bibinfo {author} {\bibfnamefont {H.}~\bibnamefont {Ji}}, \emph {et~al.},\ }\bibfield  {title} {\bibinfo {title} {Single-pixel p-graded-n junction spectrometers},\ }\href@noop {} {\bibfield  {journal} {\bibinfo  {journal} {Nature Communications}\ }\textbf {\bibinfo {volume} {15}},\ \bibinfo {pages} {1773} (\bibinfo {year} {2024})}\BibitemShut {NoStop}%
\bibitem [{\citenamefont {Liang}\ \emph {et~al.}(2025)\citenamefont {Liang}, \citenamefont {Nan}, \citenamefont {Cai}, \citenamefont {Tan}, \citenamefont {Zheng}, \citenamefont {Lu}, \citenamefont {Huang}, \citenamefont {Yan}, \citenamefont {Lei}, \citenamefont {Wen} \emph {et~al.}}]{20}%
  \BibitemOpen
  \bibfield  {author} {\bibinfo {author} {\bibfnamefont {W.}~\bibnamefont {Liang}}, \bibinfo {author} {\bibfnamefont {X.}~\bibnamefont {Nan}}, \bibinfo {author} {\bibfnamefont {W.}~\bibnamefont {Cai}}, \bibinfo {author} {\bibfnamefont {N.}~\bibnamefont {Tan}}, \bibinfo {author} {\bibfnamefont {Q.}~\bibnamefont {Zheng}}, \bibinfo {author} {\bibfnamefont {Y.}~\bibnamefont {Lu}}, \bibinfo {author} {\bibfnamefont {Y.}~\bibnamefont {Huang}}, \bibinfo {author} {\bibfnamefont {J.}~\bibnamefont {Yan}}, \bibinfo {author} {\bibfnamefont {D.}~\bibnamefont {Lei}}, \bibinfo {author} {\bibfnamefont {L.}~\bibnamefont {Wen}}, \emph {et~al.},\ }\bibfield  {title} {\bibinfo {title} {Single-pixel infrared miniaturized spectrometer enabled by ultra-broadband reconfigurable photodetection},\ }\href@noop {} {\bibfield  {journal} {\bibinfo  {journal} {Advanced Science}\ ,\ \bibinfo {pages} {2500830}} (\bibinfo {year} {2025})}\BibitemShut {NoStop}%
\bibitem [{\citenamefont {Deng}\ \emph {et~al.}(2022)\citenamefont {Deng}, \citenamefont {Zheng}, \citenamefont {Li}, \citenamefont {Zhou}, \citenamefont {Chen}, \citenamefont {Zhang}, \citenamefont {Lu}, \citenamefont {Wang}, \citenamefont {You}, \citenamefont {Li} \emph {et~al.}}]{21}%
  \BibitemOpen
  \bibfield  {author} {\bibinfo {author} {\bibfnamefont {W.}~\bibnamefont {Deng}}, \bibinfo {author} {\bibfnamefont {Z.}~\bibnamefont {Zheng}}, \bibinfo {author} {\bibfnamefont {J.}~\bibnamefont {Li}}, \bibinfo {author} {\bibfnamefont {R.}~\bibnamefont {Zhou}}, \bibinfo {author} {\bibfnamefont {X.}~\bibnamefont {Chen}}, \bibinfo {author} {\bibfnamefont {D.}~\bibnamefont {Zhang}}, \bibinfo {author} {\bibfnamefont {Y.}~\bibnamefont {Lu}}, \bibinfo {author} {\bibfnamefont {C.}~\bibnamefont {Wang}}, \bibinfo {author} {\bibfnamefont {C.}~\bibnamefont {You}}, \bibinfo {author} {\bibfnamefont {S.}~\bibnamefont {Li}}, \emph {et~al.},\ }\bibfield  {title} {\bibinfo {title} {Electrically tunable two-dimensional heterojunctions for miniaturized near-infrared spectrometers},\ }\href@noop {} {\bibfield  {journal} {\bibinfo  {journal} {Nature communications}\ }\textbf {\bibinfo {volume} {13}},\ \bibinfo {pages} {4627} (\bibinfo {year} {2022})}\BibitemShut {NoStop}%
\bibitem [{\citenamefont {Yuan}\ \emph {et~al.}(2021)\citenamefont {Yuan}, \citenamefont {Naveh}, \citenamefont {Watanabe}, \citenamefont {Taniguchi},\ and\ \citenamefont {Xia}}]{22}%
  \BibitemOpen
  \bibfield  {author} {\bibinfo {author} {\bibfnamefont {S.}~\bibnamefont {Yuan}}, \bibinfo {author} {\bibfnamefont {D.}~\bibnamefont {Naveh}}, \bibinfo {author} {\bibfnamefont {K.}~\bibnamefont {Watanabe}}, \bibinfo {author} {\bibfnamefont {T.}~\bibnamefont {Taniguchi}},\ and\ \bibinfo {author} {\bibfnamefont {F.}~\bibnamefont {Xia}},\ }\bibfield  {title} {\bibinfo {title} {A wavelength-scale black phosphorus spectrometer},\ }\href@noop {} {\bibfield  {journal} {\bibinfo  {journal} {Nature Photonics}\ }\textbf {\bibinfo {volume} {15}},\ \bibinfo {pages} {601} (\bibinfo {year} {2021})}\BibitemShut {NoStop}%
\bibitem [{\citenamefont {Yoon}\ \emph {et~al.}(2022)\citenamefont {Yoon}, \citenamefont {Fernandez}, \citenamefont {Nigmatulin}, \citenamefont {Cai}, \citenamefont {Yang}, \citenamefont {Cui}, \citenamefont {Ahmed}, \citenamefont {Cui}, \citenamefont {Uddin}, \citenamefont {Minot} \emph {et~al.}}]{23}%
  \BibitemOpen
  \bibfield  {author} {\bibinfo {author} {\bibfnamefont {H.~H.}\ \bibnamefont {Yoon}}, \bibinfo {author} {\bibfnamefont {H.~A.}\ \bibnamefont {Fernandez}}, \bibinfo {author} {\bibfnamefont {F.}~\bibnamefont {Nigmatulin}}, \bibinfo {author} {\bibfnamefont {W.}~\bibnamefont {Cai}}, \bibinfo {author} {\bibfnamefont {Z.}~\bibnamefont {Yang}}, \bibinfo {author} {\bibfnamefont {H.}~\bibnamefont {Cui}}, \bibinfo {author} {\bibfnamefont {F.}~\bibnamefont {Ahmed}}, \bibinfo {author} {\bibfnamefont {X.}~\bibnamefont {Cui}}, \bibinfo {author} {\bibfnamefont {M.~G.}\ \bibnamefont {Uddin}}, \bibinfo {author} {\bibfnamefont {E.~D.}\ \bibnamefont {Minot}}, \emph {et~al.},\ }\bibfield  {title} {\bibinfo {title} {Miniaturized spectrometers with a tunable van der waals junction},\ }\href@noop {} {\bibfield  {journal} {\bibinfo  {journal} {Science}\ }\textbf {\bibinfo {volume} {378}},\ \bibinfo {pages} {296} (\bibinfo {year} {2022})}\BibitemShut {NoStop}%
\bibitem [{\citenamefont {Tian}\ \emph {et~al.}(2024)\citenamefont {Tian}, \citenamefont {Liu}, \citenamefont {Lu}, \citenamefont {Wang}, \citenamefont {Zheng}, \citenamefont {Song}, \citenamefont {Zhong},\ and\ \citenamefont {Wang}}]{24}%
  \BibitemOpen
  \bibfield  {author} {\bibinfo {author} {\bibfnamefont {M.}~\bibnamefont {Tian}}, \bibinfo {author} {\bibfnamefont {B.}~\bibnamefont {Liu}}, \bibinfo {author} {\bibfnamefont {Z.}~\bibnamefont {Lu}}, \bibinfo {author} {\bibfnamefont {Y.}~\bibnamefont {Wang}}, \bibinfo {author} {\bibfnamefont {Z.}~\bibnamefont {Zheng}}, \bibinfo {author} {\bibfnamefont {J.}~\bibnamefont {Song}}, \bibinfo {author} {\bibfnamefont {X.}~\bibnamefont {Zhong}},\ and\ \bibinfo {author} {\bibfnamefont {F.}~\bibnamefont {Wang}},\ }\bibfield  {title} {\bibinfo {title} {Miniaturized on-chip spectrometer enabled by electrochromic modulation},\ }\href@noop {} {\bibfield  {journal} {\bibinfo  {journal} {Light: Science \& Applications}\ }\textbf {\bibinfo {volume} {13}},\ \bibinfo {pages} {278} (\bibinfo {year} {2024})}\BibitemShut {NoStop}%
\bibitem [{\citenamefont {Zhang}\ \emph {et~al.}(2025{\natexlab{a}})\citenamefont {Zhang}, \citenamefont {Albrow-Owen}, \citenamefont {Peng}, \citenamefont {Liang}, \citenamefont {Zhang}, \citenamefont {Wang}, \citenamefont {Di}, \citenamefont {Dong}, \citenamefont {Luo}, \citenamefont {Wang} \emph {et~al.}}]{25}%
  \BibitemOpen
  \bibfield  {author} {\bibinfo {author} {\bibfnamefont {G.}~\bibnamefont {Zhang}}, \bibinfo {author} {\bibfnamefont {T.}~\bibnamefont {Albrow-Owen}}, \bibinfo {author} {\bibfnamefont {W.}~\bibnamefont {Peng}}, \bibinfo {author} {\bibfnamefont {X.}~\bibnamefont {Liang}}, \bibinfo {author} {\bibfnamefont {X.}~\bibnamefont {Zhang}}, \bibinfo {author} {\bibfnamefont {P.}~\bibnamefont {Wang}}, \bibinfo {author} {\bibfnamefont {D.}~\bibnamefont {Di}}, \bibinfo {author} {\bibfnamefont {S.}~\bibnamefont {Dong}}, \bibinfo {author} {\bibfnamefont {J.}~\bibnamefont {Luo}}, \bibinfo {author} {\bibfnamefont {G.}~\bibnamefont {Wang}}, \emph {et~al.},\ }\bibfield  {title} {\bibinfo {title} {Stress-engineered ultra-broadband spectrometers},\ }\href@noop {} {\bibfield  {journal} {\bibinfo  {journal} {Science Advances}\ }\textbf {\bibinfo {volume} {11}},\ \bibinfo {pages} {eadu4225} (\bibinfo {year} {2025}{\natexlab{a}})}\BibitemShut {NoStop}%
\bibitem [{\citenamefont {Tabatabaeian}\ \emph {et~al.}(2022)\citenamefont {Tabatabaeian}, \citenamefont {Ghasemi}, \citenamefont {Shokrieh}, \citenamefont {Marzbanrad}, \citenamefont {Baraheni},\ and\ \citenamefont {Fotouhi}}]{26}%
  \BibitemOpen
  \bibfield  {author} {\bibinfo {author} {\bibfnamefont {A.}~\bibnamefont {Tabatabaeian}}, \bibinfo {author} {\bibfnamefont {A.~R.}\ \bibnamefont {Ghasemi}}, \bibinfo {author} {\bibfnamefont {M.~M.}\ \bibnamefont {Shokrieh}}, \bibinfo {author} {\bibfnamefont {B.}~\bibnamefont {Marzbanrad}}, \bibinfo {author} {\bibfnamefont {M.}~\bibnamefont {Baraheni}},\ and\ \bibinfo {author} {\bibfnamefont {M.}~\bibnamefont {Fotouhi}},\ }\bibfield  {title} {\bibinfo {title} {Residual stress in engineering materials: a review},\ }\href@noop {} {\bibfield  {journal} {\bibinfo  {journal} {Advanced engineering materials}\ }\textbf {\bibinfo {volume} {24}},\ \bibinfo {pages} {2100786} (\bibinfo {year} {2022})}\BibitemShut {NoStop}%
\bibitem [{\citenamefont {Shokrieh}(2014)}]{27}%
  \BibitemOpen
  \bibfield  {author} {\bibinfo {author} {\bibfnamefont {M.~M.}\ \bibnamefont {Shokrieh}},\ }\href@noop {} {\emph {\bibinfo {title} {Residual stresses in composite materials, 1st ed}}}\ (\bibinfo  {publisher} {Woodhead Publishing, Limited, 2014},\ \bibinfo {year} {2014})\BibitemShut {NoStop}%
\bibitem [{\citenamefont {Li}\ \emph {et~al.}(2013)\citenamefont {Li}, \citenamefont {Yin}, \citenamefont {Jiang},\ and\ \citenamefont {Zhang}}]{28}%
  \BibitemOpen
  \bibfield  {author} {\bibinfo {author} {\bibfnamefont {C.}~\bibnamefont {Li}}, \bibinfo {author} {\bibfnamefont {W.}~\bibnamefont {Yin}}, \bibinfo {author} {\bibfnamefont {H.}~\bibnamefont {Jiang}},\ and\ \bibinfo {author} {\bibfnamefont {Y.}~\bibnamefont {Zhang}},\ }\bibfield  {title} {\bibinfo {title} {An efficient augmented lagrangian method with applications to total variation minimization},\ }\href@noop {} {\bibfield  {journal} {\bibinfo  {journal} {Computational Optimization and Applications}\ }\textbf {\bibinfo {volume} {56}},\ \bibinfo {pages} {507} (\bibinfo {year} {2013})}\BibitemShut {NoStop}%
\bibitem [{\citenamefont {Song}\ \emph {et~al.}(2024)\citenamefont {Song}, \citenamefont {Liu}, \citenamefont {Wang}, \citenamefont {Chen}, \citenamefont {Shan}, \citenamefont {Zhong}, \citenamefont {Wu},\ and\ \citenamefont {Wang}}]{29}%
  \BibitemOpen
  \bibfield  {author} {\bibinfo {author} {\bibfnamefont {J.}~\bibnamefont {Song}}, \bibinfo {author} {\bibfnamefont {B.}~\bibnamefont {Liu}}, \bibinfo {author} {\bibfnamefont {Y.}~\bibnamefont {Wang}}, \bibinfo {author} {\bibfnamefont {C.}~\bibnamefont {Chen}}, \bibinfo {author} {\bibfnamefont {X.}~\bibnamefont {Shan}}, \bibinfo {author} {\bibfnamefont {X.}~\bibnamefont {Zhong}}, \bibinfo {author} {\bibfnamefont {L.-A.}\ \bibnamefont {Wu}},\ and\ \bibinfo {author} {\bibfnamefont {F.}~\bibnamefont {Wang}},\ }\bibfield  {title} {\bibinfo {title} {Computational and dark-field ghost imaging with ultraviolet light},\ }\href@noop {} {\bibfield  {journal} {\bibinfo  {journal} {Photonics Research}\ }\textbf {\bibinfo {volume} {12}},\ \bibinfo {pages} {226} (\bibinfo {year} {2024})}\BibitemShut {NoStop}%
\bibitem [{\citenamefont {Zheng}\ \emph {et~al.}(2024)\citenamefont {Zheng}, \citenamefont {Liu}, \citenamefont {Song}, \citenamefont {Ding}, \citenamefont {Zhong}, \citenamefont {Chang}, \citenamefont {Wu}, \citenamefont {McGloin},\ and\ \citenamefont {Wang}}]{30}%
  \BibitemOpen
  \bibfield  {author} {\bibinfo {author} {\bibfnamefont {Z.}~\bibnamefont {Zheng}}, \bibinfo {author} {\bibfnamefont {B.}~\bibnamefont {Liu}}, \bibinfo {author} {\bibfnamefont {J.}~\bibnamefont {Song}}, \bibinfo {author} {\bibfnamefont {L.}~\bibnamefont {Ding}}, \bibinfo {author} {\bibfnamefont {X.}~\bibnamefont {Zhong}}, \bibinfo {author} {\bibfnamefont {L.}~\bibnamefont {Chang}}, \bibinfo {author} {\bibfnamefont {X.}~\bibnamefont {Wu}}, \bibinfo {author} {\bibfnamefont {D.}~\bibnamefont {McGloin}},\ and\ \bibinfo {author} {\bibfnamefont {F.}~\bibnamefont {Wang}},\ }\bibfield  {title} {\bibinfo {title} {Temporal compressive edge imaging enabled by a lensless diffuser camera},\ }\href@noop {} {\bibfield  {journal} {\bibinfo  {journal} {Optics Letters}\ }\textbf {\bibinfo {volume} {49}},\ \bibinfo {pages} {3058} (\bibinfo {year} {2024})}\BibitemShut {NoStop}%
\bibitem [{\citenamefont {Hossain}\ \emph {et~al.}(2016)\citenamefont {Hossain}, \citenamefont {Canning}, \citenamefont {Cook},\ and\ \citenamefont {Jamalipour}}]{31}%
  \BibitemOpen
  \bibfield  {author} {\bibinfo {author} {\bibfnamefont {M.~A.}\ \bibnamefont {Hossain}}, \bibinfo {author} {\bibfnamefont {J.}~\bibnamefont {Canning}}, \bibinfo {author} {\bibfnamefont {K.}~\bibnamefont {Cook}},\ and\ \bibinfo {author} {\bibfnamefont {A.}~\bibnamefont {Jamalipour}},\ }\bibfield  {title} {\bibinfo {title} {Optical fiber smartphone spectrometer},\ }\href@noop {} {\bibfield  {journal} {\bibinfo  {journal} {Optics letters}\ }\textbf {\bibinfo {volume} {41}},\ \bibinfo {pages} {2237} (\bibinfo {year} {2016})}\BibitemShut {NoStop}%
\bibitem [{\citenamefont {Leary}\ \emph {et~al.}(2021)\citenamefont {Leary}, \citenamefont {Crocombe},\ and\ \citenamefont {Kammrath}}]{32}%
  \BibitemOpen
  \bibfield  {author} {\bibinfo {author} {\bibfnamefont {P.~E.}\ \bibnamefont {Leary}}, \bibinfo {author} {\bibfnamefont {R.~A.}\ \bibnamefont {Crocombe}},\ and\ \bibinfo {author} {\bibfnamefont {B.~W.}\ \bibnamefont {Kammrath}},\ }\bibfield  {title} {\bibinfo {title} {Introduction to portable spectroscopy},\ }\href@noop {} {\bibfield  {journal} {\bibinfo  {journal} {Portable Spectroscopy and Spectrometry}\ ,\ \bibinfo {pages} {1}} (\bibinfo {year} {2021})}\BibitemShut {NoStop}%
\bibitem [{\citenamefont {Tan}\ \emph {et~al.}(2023)\citenamefont {Tan}, \citenamefont {Li},\ and\ \citenamefont {Crozier}}]{33}%
  \BibitemOpen
  \bibfield  {author} {\bibinfo {author} {\bibfnamefont {H.}~\bibnamefont {Tan}}, \bibinfo {author} {\bibfnamefont {B.}~\bibnamefont {Li}},\ and\ \bibinfo {author} {\bibfnamefont {K.~B.}\ \bibnamefont {Crozier}},\ }\bibfield  {title} {\bibinfo {title} {Optical fiber speckle spectrometer based on reversed-lens smartphone microscope},\ }\href@noop {} {\bibfield  {journal} {\bibinfo  {journal} {Scientific Reports}\ }\textbf {\bibinfo {volume} {13}},\ \bibinfo {pages} {12958} (\bibinfo {year} {2023})}\BibitemShut {NoStop}%
\bibitem [{\citenamefont {Redding}\ \emph {et~al.}(2013)\citenamefont {Redding}, \citenamefont {Liew}, \citenamefont {Sarma},\ and\ \citenamefont {Cao}}]{34}%
  \BibitemOpen
  \bibfield  {author} {\bibinfo {author} {\bibfnamefont {B.}~\bibnamefont {Redding}}, \bibinfo {author} {\bibfnamefont {S.~F.}\ \bibnamefont {Liew}}, \bibinfo {author} {\bibfnamefont {R.}~\bibnamefont {Sarma}},\ and\ \bibinfo {author} {\bibfnamefont {H.}~\bibnamefont {Cao}},\ }\bibfield  {title} {\bibinfo {title} {Compact spectrometer based on a disordered photonic chip},\ }\href@noop {} {\bibfield  {journal} {\bibinfo  {journal} {Nature Photonics}\ }\textbf {\bibinfo {volume} {7}},\ \bibinfo {pages} {746} (\bibinfo {year} {2013})}\BibitemShut {NoStop}%
\bibitem [{\citenamefont {Chen}\ \emph {et~al.}(2024)\citenamefont {Chen}, \citenamefont {Gu},\ and\ \citenamefont {Liu}}]{35}%
  \BibitemOpen
  \bibfield  {author} {\bibinfo {author} {\bibfnamefont {C.}~\bibnamefont {Chen}}, \bibinfo {author} {\bibfnamefont {H.}~\bibnamefont {Gu}},\ and\ \bibinfo {author} {\bibfnamefont {S.}~\bibnamefont {Liu}},\ }\bibfield  {title} {\bibinfo {title} {Ultra-simplified diffraction-based computational spectrometer},\ }\href@noop {} {\bibfield  {journal} {\bibinfo  {journal} {Light: Science \& Applications}\ }\textbf {\bibinfo {volume} {13}},\ \bibinfo {pages} {9} (\bibinfo {year} {2024})}\BibitemShut {NoStop}%
\bibitem [{\citenamefont {Sun}\ \emph {et~al.}(2022)\citenamefont {Sun}, \citenamefont {Falak}, \citenamefont {Vettenburg}, \citenamefont {Lee}, \citenamefont {Phillips}, \citenamefont {Brambilla},\ and\ \citenamefont {Beresna}}]{36}%
  \BibitemOpen
  \bibfield  {author} {\bibinfo {author} {\bibfnamefont {Q.}~\bibnamefont {Sun}}, \bibinfo {author} {\bibfnamefont {P.}~\bibnamefont {Falak}}, \bibinfo {author} {\bibfnamefont {T.}~\bibnamefont {Vettenburg}}, \bibinfo {author} {\bibfnamefont {T.}~\bibnamefont {Lee}}, \bibinfo {author} {\bibfnamefont {D.~B.}\ \bibnamefont {Phillips}}, \bibinfo {author} {\bibfnamefont {G.}~\bibnamefont {Brambilla}},\ and\ \bibinfo {author} {\bibfnamefont {M.}~\bibnamefont {Beresna}},\ }\bibfield  {title} {\bibinfo {title} {Compact nano-void spectrometer based on a stable engineered scattering system},\ }\href@noop {} {\bibfield  {journal} {\bibinfo  {journal} {Photonics Research}\ }\textbf {\bibinfo {volume} {10}},\ \bibinfo {pages} {2328} (\bibinfo {year} {2022})}\BibitemShut {NoStop}%
\bibitem [{\citenamefont {Fan}\ \emph {et~al.}(2024)\citenamefont {Fan}, \citenamefont {Huang}, \citenamefont {Zhu}, \citenamefont {Liu}, \citenamefont {Jin}, \citenamefont {Guo}, \citenamefont {An}, \citenamefont {Kivshar}, \citenamefont {Qiu},\ and\ \citenamefont {Li}}]{37}%
  \BibitemOpen
  \bibfield  {author} {\bibinfo {author} {\bibfnamefont {Y.}~\bibnamefont {Fan}}, \bibinfo {author} {\bibfnamefont {W.}~\bibnamefont {Huang}}, \bibinfo {author} {\bibfnamefont {F.}~\bibnamefont {Zhu}}, \bibinfo {author} {\bibfnamefont {X.}~\bibnamefont {Liu}}, \bibinfo {author} {\bibfnamefont {C.}~\bibnamefont {Jin}}, \bibinfo {author} {\bibfnamefont {C.}~\bibnamefont {Guo}}, \bibinfo {author} {\bibfnamefont {Y.}~\bibnamefont {An}}, \bibinfo {author} {\bibfnamefont {Y.}~\bibnamefont {Kivshar}}, \bibinfo {author} {\bibfnamefont {C.-W.}\ \bibnamefont {Qiu}},\ and\ \bibinfo {author} {\bibfnamefont {W.}~\bibnamefont {Li}},\ }\bibfield  {title} {\bibinfo {title} {Dispersion-assisted high-dimensional photodetector},\ }\href@noop {} {\bibfield  {journal} {\bibinfo  {journal} {Nature}\ }\textbf {\bibinfo {volume} {630}},\ \bibinfo {pages} {77} (\bibinfo {year} {2024})}\BibitemShut {NoStop}%
\bibitem [{\citenamefont {Lin}\ \emph {et~al.}(2023)\citenamefont {Lin}, \citenamefont {Huang}, \citenamefont {Lin},\ and\ \citenamefont {Wu}}]{38}%
  \BibitemOpen
  \bibfield  {author} {\bibinfo {author} {\bibfnamefont {C.-H.}\ \bibnamefont {Lin}}, \bibinfo {author} {\bibfnamefont {S.-H.}\ \bibnamefont {Huang}}, \bibinfo {author} {\bibfnamefont {T.-H.}\ \bibnamefont {Lin}},\ and\ \bibinfo {author} {\bibfnamefont {P.~C.}\ \bibnamefont {Wu}},\ }\bibfield  {title} {\bibinfo {title} {Metasurface-empowered snapshot hyperspectral imaging with convex/deep (code) small-data learning theory},\ }\href@noop {} {\bibfield  {journal} {\bibinfo  {journal} {Nature communications}\ }\textbf {\bibinfo {volume} {14}},\ \bibinfo {pages} {6979} (\bibinfo {year} {2023})}\BibitemShut {NoStop}%
\bibitem [{\citenamefont {Liu}\ \emph {et~al.}(2021)\citenamefont {Liu}, \citenamefont {Wang}, \citenamefont {Chen}, \citenamefont {Dong},\ and\ \citenamefont {McGloin}}]{39}%
  \BibitemOpen
  \bibfield  {author} {\bibinfo {author} {\bibfnamefont {B.}~\bibnamefont {Liu}}, \bibinfo {author} {\bibfnamefont {F.}~\bibnamefont {Wang}}, \bibinfo {author} {\bibfnamefont {C.}~\bibnamefont {Chen}}, \bibinfo {author} {\bibfnamefont {F.}~\bibnamefont {Dong}},\ and\ \bibinfo {author} {\bibfnamefont {D.}~\bibnamefont {McGloin}},\ }\bibfield  {title} {\bibinfo {title} {Self-evolving ghost imaging},\ }\href@noop {} {\bibfield  {journal} {\bibinfo  {journal} {Optica}\ }\textbf {\bibinfo {volume} {8}},\ \bibinfo {pages} {1340} (\bibinfo {year} {2021})}\BibitemShut {NoStop}%
\bibitem [{\citenamefont {Wang}\ \emph {et~al.}(2023)\citenamefont {Wang}, \citenamefont {Liu}, \citenamefont {Song}, \citenamefont {Wang}, \citenamefont {Shan}, \citenamefont {Zhong},\ and\ \citenamefont {Wang}}]{40}%
  \BibitemOpen
  \bibfield  {author} {\bibinfo {author} {\bibfnamefont {D.}~\bibnamefont {Wang}}, \bibinfo {author} {\bibfnamefont {B.}~\bibnamefont {Liu}}, \bibinfo {author} {\bibfnamefont {J.}~\bibnamefont {Song}}, \bibinfo {author} {\bibfnamefont {Y.}~\bibnamefont {Wang}}, \bibinfo {author} {\bibfnamefont {X.}~\bibnamefont {Shan}}, \bibinfo {author} {\bibfnamefont {X.}~\bibnamefont {Zhong}},\ and\ \bibinfo {author} {\bibfnamefont {F.}~\bibnamefont {Wang}},\ }\bibfield  {title} {\bibinfo {title} {Dual-mode adaptive-svd ghost imaging},\ }\href@noop {} {\bibfield  {journal} {\bibinfo  {journal} {Optics Express}\ }\textbf {\bibinfo {volume} {31}},\ \bibinfo {pages} {14225} (\bibinfo {year} {2023})}\BibitemShut {NoStop}%
\bibitem [{\citenamefont {Zheng}\ \emph {et~al.}(2021)\citenamefont {Zheng}, \citenamefont {Liu}, \citenamefont {Meng}, \citenamefont {Qiao}, \citenamefont {Tong}, \citenamefont {Yang}, \citenamefont {Han},\ and\ \citenamefont {Yuan}}]{41}%
  \BibitemOpen
  \bibfield  {author} {\bibinfo {author} {\bibfnamefont {S.}~\bibnamefont {Zheng}}, \bibinfo {author} {\bibfnamefont {Y.}~\bibnamefont {Liu}}, \bibinfo {author} {\bibfnamefont {Z.}~\bibnamefont {Meng}}, \bibinfo {author} {\bibfnamefont {M.}~\bibnamefont {Qiao}}, \bibinfo {author} {\bibfnamefont {Z.}~\bibnamefont {Tong}}, \bibinfo {author} {\bibfnamefont {X.}~\bibnamefont {Yang}}, \bibinfo {author} {\bibfnamefont {S.}~\bibnamefont {Han}},\ and\ \bibinfo {author} {\bibfnamefont {X.}~\bibnamefont {Yuan}},\ }\bibfield  {title} {\bibinfo {title} {Deep plug-and-play priors for spectral snapshot compressive imaging},\ }\href@noop {} {\bibfield  {journal} {\bibinfo  {journal} {Photonics Research}\ }\textbf {\bibinfo {volume} {9}},\ \bibinfo {pages} {B18} (\bibinfo {year} {2021})}\BibitemShut {NoStop}%
\bibitem [{\citenamefont {Zhang}\ \emph {et~al.}(2025{\natexlab{b}})\citenamefont {Zhang}, \citenamefont {Liu}, \citenamefont {Wang}, \citenamefont {Ma}, \citenamefont {Zheng}, \citenamefont {Liu}, \citenamefont {Huang}, \citenamefont {Wang},\ and\ \citenamefont {Song}}]{42}%
  \BibitemOpen
  \bibfield  {author} {\bibinfo {author} {\bibfnamefont {C.}~\bibnamefont {Zhang}}, \bibinfo {author} {\bibfnamefont {X.}~\bibnamefont {Liu}}, \bibinfo {author} {\bibfnamefont {L.}~\bibnamefont {Wang}}, \bibinfo {author} {\bibfnamefont {S.}~\bibnamefont {Ma}}, \bibinfo {author} {\bibfnamefont {Y.}~\bibnamefont {Zheng}}, \bibinfo {author} {\bibfnamefont {Y.}~\bibnamefont {Liu}}, \bibinfo {author} {\bibfnamefont {H.}~\bibnamefont {Huang}}, \bibinfo {author} {\bibfnamefont {Y.}~\bibnamefont {Wang}},\ and\ \bibinfo {author} {\bibfnamefont {W.}~\bibnamefont {Song}},\ }\bibfield  {title} {\bibinfo {title} {Lensless efficient snapshot hyperspectral imaging using dynamic phase modulation},\ }\href@noop {} {\bibfield  {journal} {\bibinfo  {journal} {Photonics Research}\ }\textbf {\bibinfo {volume} {13}},\ \bibinfo {pages} {511} (\bibinfo {year} {2025}{\natexlab{b}})}\BibitemShut {NoStop}%
\end{thebibliography}%

\end{document}